%% file: Main.tex
\documentclass[journal]{IEEEtran}

\input{settings_custom}
\input{acronyms.tex}

\begin{document}

\title{\huge{%
    Harmonic Stability Analysis of Microgrids with Converter-Interfaced Distributed Energy Resources,\\
	Part I: Modelling and Theoretical Foundations
}}

\author{%
	Johanna~Kristin~Maria~Becker,~\IEEEmembership{Student Member,~IEEE},
	Andreas~Martin~Kettner,~\IEEEmembership{Member,~IEEE},\\
	and~Mario~Paolone,~\IEEEmembership{Fellow,~IEEE}%
	\thanks{J. Becker and M. Paolone are with the Distributed Electrical Systems Laboratory at the {\'E}cole Polytechnique F{\'e}d{\'e}rale de Lausanne (EPFL) in CH-1015 Lausanne, Switzerland (E-mail: \{johanna.becker, mario.paolone\}@epfl.ch).}%
	\thanks{A. Kettner is with PSI NEPLAN AG, 8700 Küsnacht, Switzerland (E-mail: andreas.kettner@neplan.ch).}%
	\thanks{This work was funded by the Schweizerischer Nationalfonds (SNF, Swiss National Science Foundation) via the National Research Programme NRP~70 ``Energy Turnaround'' (projects nr. 173661 and 197060).
    }%
}

\maketitle






\input{Sections/Abstract_Keywords}



\input{Sections/Introduction}

\input{Sections/Literature_Review}

\input{Sections/Resource_Model}

\input{Sections/Grid_Model}

\input{Sections/System_Model}

\input{Sections/Operators}

\input{Sections/Conclusion}





\bibliographystyle{IEEEtran}
\bibliography{Bibliography}

\end{document}

%% file: settings_custom.tex
\usepackage{cite} 

\usepackage[caption=false,font=footnotesize]{subfig}	
\usepackage{graphicx}

\usepackage{algorithm}
\usepackage{algorithmicx}
\usepackage{algpseudocode}

\usepackage{amsmath,amsfonts,amssymb,amsthm}
\usepackage{mathtools}
\usepackage{mathrsfs}	
\usepackage{subdepth}	

\usepackage{array}	
\usepackage{tabularx}   

\usepackage{hyperref}
\usepackage[sort,compress,capitalise]{cleveref}	
\usepackage{url}

\usepackage[dvipsnames]{xcolor} 
\usepackage{xspace} 

\usepackage{tikz}
\usetikzlibrary{babel}  
\usepackage[straightvoltages,nooldvoltagedirection]{circuitikz}

\usepackage{textcomp}
\usepackage{csquotes}

\usepackage{graphicx}
\usepackage{float}

\usepackage{algorithm}
\usepackage{algorithmicx}
\usepackage{algpseudocode}

\usepackage{amsmath,amsfonts,amssymb,amsthm}
\usepackage{mathtools}
\usepackage{mathrsfs}	
\usepackage{subdepth}	
\usepackage{nicefrac}

\usepackage{array}	
\usepackage{arydshln}   
\usepackage{mathtools}
\usepackage{tabularx}   
\usepackage{tablefootnote}

\usepackage{hyperref}
\usepackage[sort,compress,capitalise]{cleveref}	
\usepackage{url}

\usepackage{xspace} 

\usepackage{tikz}
\usetikzlibrary{fit}
\usetikzlibrary{matrix,decorations.pathreplacing, calc, positioning}
\usetikzlibrary{shapes.symbols}
\usetikzlibrary{babel}  
\usepackage[straightvoltages,nooldvoltagedirection]{circuitikz}
\usetikzlibrary{calc,shapes.geometric}
\usepackage{scalerel}
\usetikzlibrary{arrows.meta}
\usepackage{circuitikz}

\usepackage{nomencl}

{
\theoremstyle{plain}
\newtheorem{Hypothesis}{Hypothesis}

\newtheorem{Definition}{Definition}
}

{
\theoremstyle{definition}

}

\crefname{Hypothesis}{Hyp.}{Hyps.}
\Crefname{Hypothesis}{Hyp.}{Hyps.}

\crefname{Lemma}{Lemma}{Lemmata}
\Crefname{Lemma}{Lemma}{Lemmata}

\crefname{Definition}{Def.}{Defs.}
\Crefname{Definition}{Def.}{Defs.}

\definecolor{color1}{rgb}{0.510,0.129,0.549}
\definecolor{color2}{rgb}{0.000,0.259,0.549}
\definecolor{color3}{rgb}{0.000,0.451,0.329}
\definecolor{color4}{rgb}{0.961,0.608,0.137}
\definecolor{color5}{rgb}{1.000,0.592,0.592}


%% file: acronyms.tex
\newcommand{\CIGRE}{CIGR{\'E}\xspace}

\newcommand{\DFT}[1][]{DFT\xspace}  
\nomenclature{\DFT}{Discrete Fourier Transform}

\newcommand{\ADC}[1][]{ADC#1\xspace}  
\nomenclature{\ADC}{Analog-to-Digital Converter}
\newcommand{\DAC}[1][]{DAC#1\xspace}  
\nomenclature{\DAC}{Digital-to-Analog Converter}

\newcommand{\LPF}[1][]{LPF#1\xspace}  
\nomenclature{\LPF}{Low-Pass Filter}
\newcommand{\RMS}{RMS\xspace}   
\nomenclature{\RMS}{Root-Mean-Square}

\newcommand{\ctrlPID}{\textup{PID}\xspace}
\nomenclature{\ctrlPID}{Proportional-Integral-Derivative}
\newcommand{\ctrlPR}{\textup{PR}\xspace}
\nomenclature{\ctrlPR}{Proportional-Resonant}

\newcommand{\ctrlFB}{\textup{\texttt{FB}}\xspace}   
\nomenclature{\ctrlFB}{Feed-Back}
\newcommand{\ctrlFF}{\textup{\texttt{FF}}\xspace}   
\nomenclature{\ctrlFF}{Feed-Forward}
 
\newcommand{\ctrlFT}{\textup{\texttt{FT}}\xspace}   
\nomenclature{\ctrlFT}{Feed-Through}

\newcommand{\MIMO}{MIMO\xspace}	
\nomenclature{\MIMO}{Multi-Input Multi-Output}
\newcommand{\LTI}{LTI\xspace}	
\nomenclature{\LTI}{Linear Time-Invariant}
\newcommand{\LTP}{LTP\xspace}	
\nomenclature{\LTP}{Linear Time-Periodic}
\newcommand{\EHD}{EHD\xspace}	
\nomenclature{\EHD}{Extended Harmonic Domain}
\newcommand{\DPM}{DP\xspace}	
\nomenclature{\DPM}{Dynamic Phasor }
\newcommand{\EMP}{EMP\xspace}	
\nomenclature{\EMP}{Exponentially Modulated Time-Periodic}

\newcommand{\TE}[1][]{\textup{TE#1}\xspace}   
\nomenclature{\TE}{Th{\'e}venin Equivalent}
\newcommand{\NE}[1][]{\textup{NE#1}\xspace}   
\nomenclature{\NE}{Norton Equivalent}

\newcommand{\DAE}[1][]{DAE#1\xspace}		
\nomenclature{\DAE}{Differential-Algebraic Equation}

\newcommand{\MNA}{MNA\xspace}		
\nomenclature{\MNA}{Modified Nodal Analysis}
\newcommand{\MANA}{MANA\xspace}		
\nomenclature{\MANA}{Modified Augmented Nodal Analysis}

\newcommand{\AC}{AC\xspace} 
\newcommand{\DC}{DC\xspace}	

\newcommand{\ADN}[1][]{ADN#1\xspace}	
\nomenclature{\ADN}{Active Distribution Network}
\newcommand{\DER}[1][]{DER#1\xspace}    
\newcommand{\CIDER}[1][]{CI\DER[#1]}    
\nomenclature{\CIDER}{Converter-Interfaced Distributed Energy Resource}
\newcommand{\NIC}[1][]{NIC#1\xspace}         
\nomenclature{\NIC}{Network-Interfacing Converter}

\newcommand{\TDS}{TDS\xspace}	
\nomenclature{\TDS}{Time-Domain Simulation}
\newcommand{\HA}{HA\xspace}		
\nomenclature{\HA}{Harmonic Analysis}
\newcommand{\DHA}{D\HA}			
\nomenclature{\DHA}{Direct Harmonic Analysis}
\newcommand{\IHA}{I\HA}			
\nomenclature{\IHA}{Iterative Harmonic Analysis}
\newcommand{\HDR}[1][]{HDR#1\xspace}	
\nomenclature{\HDR}{Harmonic Domain Response}
\newcommand{\HPF}{HPF\xspace}	
\nomenclature{\HPF}{Harmonic Power Flow}
\newcommand{\HSS}{HSS\xspace}	
\nomenclature{\HSS}{Harmonic State-Space}
\newcommand{\HTF}{HTF\xspace}	
\nomenclature{\HTF}{Harmonic Transfer Function}
\newcommand{\HSA}{HSA\xspace}	
\nomenclature{\HSA}{Harmonic Stability Assessment}

\newcommand{\DI}{DI\xspace}	
\nomenclature{\DI}{Design Invariant}
\newcommand{\CDI}{CDI\xspace}	
\nomenclature{\CDI}{Control-Design Invariant}
\newcommand{\CDV}{CDV\xspace}	
\nomenclature{\CDV}{Control-Design Variant}
\newcommand{\LAP}{LAP\xspace}	
\nomenclature{\LAP}{Linear Assignment Problem}

\newcommand{\EMTP}{EMTP\xspace}		
\nomenclature{\EMTP}{Electromagnetic Transients Program}
\newcommand{\SPICE}{SPICE\xspace}	
\nomenclature{\SPICE}{Simulation Program with Integrated Circuit Emphasis}

\newcommand{\PLL}[1][]{PLL#1\xspace}		
\nomenclature{\PLL}{Phase-Locked Loop}
\newcommand{\PWM}{PWM\xspace}	
\nomenclature{\PWM}{Pulse-Width Modulator}

\newcommand{\KPI}[1][]{KPI#1\xspace}    
\nomenclature{\KPI}{Key Performance Indicator}
\newcommand{\THD}[1][]{THD#1\xspace}    
\nomenclature{\THD}{Total Harmonic Distortion}
\newcommand{\DFS}[1][]{DFS#1\xspace}	
\nomenclature{\DFS}{Double Fourier Series}

\newcommand{\Exp}[1]{\exp\left(#1\right)}

\newcommand{\Set}[1]{\mathcal{#1}}

\newcommand{\diag}{\operatorname{diag}}

\newcommand{\col}{\operatorname{col}}

\newcommand{\grd}{\gamma}
\newcommand{\pwr}{\pi}
\newcommand{\ctrl}{\kappa}

\newcommand{\refr}{\rho}
\newcommand{\spt}{\sigma}
\newcommand{\opt}{o}

\newcommand{\Y}{\mathbf{Y}}     

\newcommand{\nodes}{\Set{N}}
\newcommand{\branches}{\Set{L}}
\newcommand{\shunts}{\Set{T}}

\newcommand{\Frm}{\Set{S}}
\newcommand{\frm}{s}
\newcommand{\Flw}{\Set{R}}
\newcommand{\flw}{r}

\newcommand{\Prc}{\mathcal{P}}
\newcommand{\Grd}{\mathcal{G}}

\newcommand{\Rsc}{\mathcal{Q}}
\newcommand{\rsc}{q}

\newcommand{\harmonics}{\Set{H}}

\newcommand{\VT}{\mathbf{v}}
\newcommand{\IT}{\mathbf{i}}
\newcommand{\XT}{\mathbf{x}}

\newcommand{\WT}{\mathbf{w}}

\newcommand{\YT}{\mathbf{y}}

\newcommand{\CP}{\mathbf{C}}

\newcommand{\RP}{\mathbf{R}}
\newcommand{\LP}{\mathbf{L}}



%% file: Sections/Abstract_Keywords.tex

\begin{abstract}
    This paper proposes a method for the \emph{Harmonic Stability Assessment} (\HSA) of power systems with a high share of \emph{Converter-Interfaced Distributed Energy Resources} (\CIDER[s]).
	To this end, the \emph{Harmonic State-Space} (\HSS) model of a generic power system is formulated by combining the \HSS models of the resources and the grid in closed-loop configuration.
	The \HSS model of the resources is obtained from the \emph{Linear Time-Periodic} (\LTP) models of the \CIDER components transformed to frequency domain using Fourier theory and Toeplitz matrices.
    Notably, the \HSS of a \CIDER is capable of representing the coupling between harmonic frequencies in detail.
	The \HSS model of the grid is derived from the dynamic equations of the individual branch and shunt elements.
	The system matrix of the \HSS models on power-system or resource level is employed for eigenvalue analysis in the context of \HSA.
	A sensitivity analysis of the eigenvalue loci w.r.t. changes in model parameters, and a classification of eigenvalues into control-design variant, control-design invariant, and design invariant eigenvalues is proposed.
    A case of harmonic instability is identified by the \HSA and validated via \emph{Time-Domain Simulations} (\TDS) in Simulink.
\end{abstract}



%% file: Sections/Introduction.tex
\section{Introduction}
\label{sec:intro}

%
%
%



\IEEEPARstart{P}{ower} distribution systems are increasingly integrating distributed energy resources typically connected to the grid via power electronic converters.
The extensive integration of such \emph{Converter-Interfaced Distributed Energy Resources} (\CIDER[s]) can potentially threaten power system stability.
It is therefore essential to develop analysis methods that are capable of identifying the sources of instability in such systems.



For over a century, power system analysis has focused on investigating the fundamental frequency component \cite{Bk:1994:Kundur}.
Yet, power systems are intrinsically complex, nonlinear systems that are characterized by time-varying signals featuring a broad spectrum of frequencies~\cite{Jrn:Paolone:2020}.
This aspect becomes significantly more pronounced in power systems with a high share of \CIDER[s], where interactions between frequencies occur. 
Recent findings show that modern power systems tend to reach a so-called \emph{harmonic steady state}, as opposed to the traditionally considered (fundamental tone) \emph{sinusoidal steady state}.

Therefore, the traditional concepts of power system stability are being reevaluated.
More precisely, several standardization committees have been focusing on the classification, modelling, and examination of stability issues in distribution grids (e.g., \cite{Rep:PSE:SA:2018:Canizares,Rep:PSE:SA:2020:Hatziargyriou}).
Given the widespread integration of \CIDER[s], \emph{converter} a.k.a \emph{harmonic stability} has become particularly critical \cite{Rep:PSE:SA:2018:Canizares}.
In such cases, the presence of \CIDER[s] in the system induce unstable oscillations at harmonic frequencies (e.g., \cite{Jrn:PSE:PEC:2004:Enslin}) due to interactions between \AC/\DC converters and their components, as well as the collective interactions of various \CIDER[s] through the electrical grid.


For the \emph{Harmonic Stability Assessment} (\HSA) of power systems with a high share of \CIDER[s], the system and its components are typically described by \emph{Linear Time-Periodic} (\LTP) models \cite{Jrn:Wang:2018}.
These \LTP models are converted to the frequency domain using Fourier and Toeplitz theory~\cite{Ths:CSE:LTP:1991:Wereley}, leading to the so called \emph{Harmonic State Space} (\HSS) models.
Such \HSS models are capable of representing the coupling between harmonic frequencies, that is overlooked by the traditionally used \emph{Linear Time-Invariant} (\LTI) models.
Based on the \HSS models, the system's small-signal harmonic stability at a particular operating point can be examined using Nyquist plots or eigenvalue analysis~\cite{Rep:PSE:SA:2004:Kundur}.



This two-part paper proposes a method for the \HSA of generic power systems.
Specifically, the \HSS model of the system, that effectively represents the coupling between harmonic frequencies, can be employed for \HSA through eigenvalue analysis.
The main contributions of Part~I are as follows:
\begin{itemize}
    \item
        Derivation of the \HSS models of individual \CIDER[s], the grid, as well as whole power systems based on the theory provided in \cite{jrn:2020:kettner-becker:HPF-1,jrn:2020:kettner-becker:HPF-2}.
    \item 
        Derivation of operators for (i) the sensitivity analysis of eigenvalue loci w.r.t.~control parameter variations, and (ii) the classification of eigenvalues into \emph{control-design variant}, \emph{control-design invariant} and \emph{design invariant}. 
\end{itemize}
The main contributions of Part~II are:
\begin{itemize}
    \item 
        Detailed discussion of the \HSA for the \CIDER[s] introduced in \cite{jrn:2020:kettner-becker:HPF-2, jrn:2022:becker} and a small yet realistic power system derived from the \CIGRE benchmark system \cite{Rep:2014:CIGRE}.
    \item  
        Identification of a harmonic instability in the adopted test system by means of the system eigenvalues, which is validated through \emph{Time-Domain Simulations} (\TDS) in Simulink.
\end{itemize}
\noindent
The remainder of this part is organised as follows.
\cref{sec:soa} provides a comprehensive literature review on existing methods for \HSA.
\cref{sec:rsc-hss} and \cref{sec:grd-hss} introduce the generic \HSS model of the individual \CIDER and the grid, respectively.
These models are combined to the \HSS model of the entire power system in \cref{sec:sys-hss}.
In \cref{sec:op-hsa} the operators used for the \HSA of generic \HSS models are introduced and an illustration of \LTP eigenvalues is given.
\cref{sec:conclusion} draws some first conclusions.

%% file: Sections/Literature_Review.tex
\section{Literature Review}
\label{sec:soa}

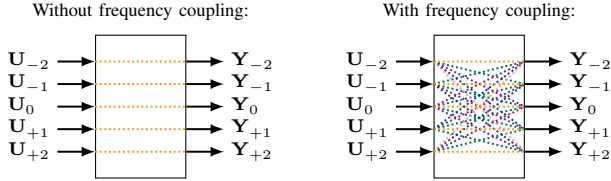
\begin{figure}[t]
	\centering
	\input{Figures/HarmonicCoupling.tex}
	\caption
	{%
		Illustration of the effect of frequency coupling in a model.
		In- and output signals are depicted through their Fourier coefficients, assuming a maximum harmonic order of two.
        The model behaviour w.r.t.~coupling of frequencies is illustrated by rectangular blocks.
		The positive and negative spectrum of the signal are taken into account.
	}
	\label{fig:mf:ltp-coupling}
\end{figure}

The presence of harmonics and the nonlinear characteristics of certain elements in modern power systems (e.g., \CIDER[s]) results in a coupling between frequencies.
This phenomenon is often overlooked in traditional power system analysis.
Therefore, a critical factor to consider when categorizing these methods is their ability to account for the effect of frequency coupling, as visualized in \cref{fig:mf:ltp-coupling}.
A model ignoring frequency coupling translates inputs to outputs of the same frequency with altered magnitude and phase, whereas models accounting for frequency coupling can produce outputs across the entire spectrum of frequencies from a single input frequency.

\subsection{Basics of Harmonic Steady State}

Operation of power systems is typically divided into a sequence of quasi-steady states \cite{Bk:PSE:SSA:1997:Arrillaga}.
Within the scope of \HSA, it is crucial to focus on generic \emph{periodic steady states} which comprise harmonic frequencies (i.e., as opposed to \emph{sinusoidal steady states}).
To reduce the computational complexity of the models, harmonic phasors, which are the Fourier coefficients of the signals at these harmonic frequencies, are utilized.
Such frequency-domain modelling approaches are particularly suitable for analyses that focus on stability around a specific harmonic steady state.
The ability to assess this specific type of instability, that stems from the presence of harmonics as well as the interaction between them, largely depends on the characteristics of the adopted models.
Hence, the following literature review is distinguishing the \HSA methods based on approaches that either represent or do not represent the frequency coupling.
Detailed discussions on this topic can be found in \cite{Jrn:Wang:2018,Jrn:2017:Kwon}. 


\subsection{Approaches without Representation of Frequency Coupling}

In sinusoidal steady state, both the grid and the resources (including their controllers) can be described using symmetrical components and the Park transform \cite{Jrn:Park:1929}.
This results in \LTI models that can be examined in frequency domain at the harmonic frequencies.
However, such a model is unable to account for frequency coupling, since it relies on transfer functions (recall \cref{fig:mf:ltp-coupling}).
While stability criteria derived from such models may be able to identify instabilities caused by a single harmonic, they would fail to detect an instability arising from the interaction of multiple harmonics (i.e., coupling).
Additionally, since asymmetries (e.g., w.r.t.~to the phase conductors) cannot be converted into time-invariant behaviour by the Park transform, this method is limited to analysing symmetrical system components.
A detailed overview of such modelling approaches is given in Chapter~4 of \cite{Bk:PSE:SSA:1997:Arrillaga}.
Specifically, the harmonic sources of the system are assumed to be decoupled from harmonic flows and the power system components (such as generators, transformers, and loads) are represented by specific impedances at each harmonic frequency.
In both \cite{Jrn:Wang:2014} and \cite{Jrn:Yoon:2016}, the system components are described using impedance models, which are based on \LTI models.
These impedances are then employed for harmonic analyses through Nyquist theory.

\subsection{Approaches with Representation of Frequency Coupling}

For the \HSA of models capturing frequency coupling, two types of modelling approaches are considered: \HSS models, which are commonly used, and \emph{Dynamic Phasors}~(\DPM).

In the \HSS approach, the \LTP model of the system is obtained via linearization around time-periodic trajectories.
This \LTP model is then transformed to the harmonic domain using Toeplitz matrices to create the \HSS model~\cite{Ths:CSE:LTP:1991:Wereley}.
In the last decades, the \HSS approach has gained significant attention in research, particularly for modelling power systems featuring a substantial amount of \CIDER[s].
Several works have advanced the understanding of harmonic coupling within the individual \CIDER[s] occurring due to control interactions as well as synchronization effects~\cite{Jrn:Kwon:2016,Jrn:Kwon:2016-1,Jrn:2022:Yang,Jrn:Golestan:2021,Cnf:Salis:2016}.

The \DPM method is a modelling technique intended for the dynamic study of power systems taking into account the effect of frequency coupling~\cite{Cnf:Stankovic:2000}.
It operates in the time domain, while integrating methods and concepts from frequency domain.
Specifically, the system variables are decomposed into harmonic phasors with time-variant Fourier coefficients.
For the \DPM models, the nonlinear system equations can either be expressed in terms of the dynamic phasors, or an \LTP model is determined via small-signal analysis.
\DPM models have been used for \TDS of power systems, as shown in \cite{Cnf:Demiray:2008,Cnf:Stankovic:2000,Jrn:DeRua:2020}.

Literature on the \HSS approach frequently uses both Nyquist and eigenvalue techniques, whereas the \DPM approach predominantly utilizes eigenvalue analysis.
The following paragraphs are structured w.r.t. to the analysis technique used for the \HSA.

\subsubsection{\HSA Employing the Nyquist Criterion}

In \cite{Cnf:Hall:1990}, a generalization of the Nyquist criterion to \HSS models, which is similar to the generalized Nyquist criterion for \LTI \emph{multi-input multi-output} systems, was first proposed.
In \cite{Jrn:Salis:2017}, the equivalent harmonic impedances of a single-phase system, comprising two \CIDER[s], are identified through small-signal current injections and subsequently used to derive the Nyquist plot of the system.
In \cite{Cnf:Kwon:2016}, the Nyquist plots for a single-phase grid-connected \CIDER are derived employing the equivalent impedance of the \CIDER and the grid obtained from the corresponding \HSS models.

\subsubsection{\HSA through Eigenvalue Analysis}

Many studies use the eigenvalues of the \HSS model or poles of the corresponding \emph{Harmonic Transfer Function} (\HTF) for stability assessment.
The theory behind this was first detailed in \cite{Ths:CSE:LTP:1991:Wereley}. 
Current research often aims at understanding the causes of harmonic instability by investigating individual \CIDER[s].
\cite{Jrn:2022:Yang} provides an in-depth analysis of stability margins in grid-following \CIDER[s] with different synchronization techniques and control approaches.
The effects of parameter variations on the stability margin are identified through eigenvalue analysis and validated in power-hardware-in-the-loop experiments.
In \cite{Jrn:Golestan:2021} a detailed study of the \HSA of a specific synchronization technique used in \CIDER[s], i.e., the so-called  multiple second-order generalized integrator is proposed.
Both \cite{Jrn:Kwon:2016-1} and \cite{Cnf:Salis:2016} examine the stability of a single-phase \CIDER connected to the grid, with the grid modelled as a \TE.
In \cite{Jrn:Kwon:2016-1}, an instability is observed by means of the poles of the \HTF, and validated through \TDS and experiments. 
\cite{Cnf:Salis:2016} examines the stability of the system through its eigenvalues and confirms the findings through \TDS.

In \cite{Jrn:Peng:2019}, a nonlinear \DPM model is derived and its small-signal model is employed for eigenvalue analysis.
For a power system including three \CIDER[s], it is shown how, upon a parameter change, the harmonic components of the active power become unstable, while the fundamental component remains in steady state.
The instability is also observed in the eigenvalues.
In another study, the eigenvalues derived from an \HSS model and a \DPM model for an individual three-phase \CIDER are compared~\cite{Jrn:DeRua:2020}. 
It is demonstrated that the eigenvalues from both methodologies coincide for the analysed \CIDER, which suggests that both approaches are equally effective in representing the coupling between harmonics and the corresponding stability margins.

\subsubsection{Modal Analysis in the context of \HSA}

When conducting an eigenvalue analysis for the \HSS system, additional insights can be gained from the associated eigenvectors.
Specifically, each entry in an eigenvector is associated with a particular eigenvalue and quantifies its impact on the respective states.
For \LTI systems, this approach is known as \emph{participation factor analysis} or \emph{modal analysis} (e.g., \cite{Bk:1994:Kundur}).
This concept is generalized from \LTI to \LTP systems (i.e., to \HSS models) in \cite{Jrn:2020:Yang}.
A detailed analysis in this respect is performed for the example of a grid-following \CIDER.



\subsection{Motivation for the Proposed Method}

In summary, \DPM and \HSS methodologies appear to be promising for \HSA due to their capability of representing the frequency-coupling effects inherent to \CIDER[s]. 
By contrast, other approaches fall short in capturing this phenomenon.

W.r.t.~modularity and ease of representation, \HSS models appear to be particularly promising. 
Extensive research has been performed on the \HSA of single \CIDER[s] or on small systems as examples.
Yet, no universal and generalised approach has been proposed being capable of analysing generic power systems while taking into account the interactions of different \CIDER[s] through the grid.
Addressing this research gap, the modelling framework that has been proposed in~\cite{jrn:2020:kettner-becker:HPF-1,jrn:2020:kettner-becker:HPF-2}, and further extended in~\cite{jrn:2022:becker}, is adapted for the \HSA based on eigenvalue analysis in this paper.

%% file: Figures/HarmonicCoupling.tex
{
	\tikzstyle{square}=[rectangle,draw=black,minimum height=1.9cm,minimum width=1.2cm,inner sep=0mm]
	\tikzstyle{UG1}=[]
	\tikzstyle{UG3}=[dashed]
	\tikzstyle{signalP}=[-latex,color2,line width = 1pt]
	\tikzstyle{signalN}=[-latex,color3,line width = 1pt]
	\tikzstyle{signalH}=[-latex,line width = 0.75pt]
	\tikzstyle{signalIn}=[densely dotted,color4,line width = 0.75pt]
	\tikzstyle{signalInC}=[densely dotted,color3,line width = 0.75pt]
	\tikzstyle{signalInC2}=[densely dotted,color1,line width = 0.75pt]
	\tikzstyle{signalInC3}=[densely dotted,color2,line width = 0.75pt]
	\tikzstyle{signalInC4}=[densely dotted,color5,line width = 0.75pt]
	
	\begin{circuitikz}
		\scriptsize
		
		\def\BlockSize{1.0}	
		\def\X{0.5}	
		\def\Y{2}	
		\def\dlX{0.5}
		\def\dlY{0.3}
		\def\dload{0.4}
		

		\node[square] (S1) at (0,0) {};
		\node (T1) at ($(S1.north)+(0,\dlY)$) {Without frequency coupling:};
		
		\draw[signalH] ($(S1.west)+(-\X,0)+(0,2*\dlY)$) to  node[at start,left]{${\mathbf{U}}_{-2}$} ($(S1.west)+(0,2*\dlY)$);
		\draw[signalH] ($(S1.west)+(-\X,0)+(0,\dlY)$) to node[at start,left]{${\mathbf{U}}_{-1}$} ($(S1.west)+(0,\dlY)$);
		\draw[signalH] ($(S1.west)+(-\X,0)$) to node[at start,left]{${\mathbf{U}}_{0\phantom{-}}$} (S1.west);
		\draw[signalH] ($(S1.west)+(-\X,0)+(0,-\dlY)$) to node[at start,left]{${\mathbf{U}}_{+1}$} ($(S1.west)+(0,-\dlY)$);
		\draw[signalH] ($(S1.west)+(-\X,0)+(0,-2*\dlY)$) to node[at start,left]{${\mathbf{U}}_{+2}$} ($(S1.west)+(0,-2*\dlY)$);
		
		\draw[signalIn] ($(S1.west)+(0,2*\dlY)$) to ($(S1.east)+(0,2*\dlY)$);
		\draw[signalIn] ($(S1.west)+(0,\dlY)$) to ($(S1.east)+(0,\dlY)$);
		\draw[signalIn] ($(S1.west)$) to (S1.east);
		\draw[signalIn] ($(S1.west)+(0,-\dlY)$) to ($(S1.east)+(0,-\dlY)$);
		\draw[signalIn] ($(S1.west)+(0,-2*\dlY)$) to ($(S1.east)+(0,-2*\dlY)$);

		\draw[signalH] ($(S1.east)+(0,2*\dlY)$) to node[at end,right]{${\mathbf{Y}}_{-2}$} ($(S1.east)+(\X,0)+(0,2*\dlY)$);
		\draw[signalH] ($(S1.east)+(0,\dlY)$) to node[at end,right]{${\mathbf{Y}}_{-1}$} ($(S1.east)+(\X,0)+(0,\dlY)$);
		\draw[signalH] (S1.east) to node[at end,right]{${\mathbf{Y}}_{0\phantom{-}}$} ($(S1.east)+\X*(1,0)$);
		\draw[signalH] ($(S1.east)+(0,-\dlY)$) to node[at end,right]{${\mathbf{Y}}_{+1}$} ($(S1.east)+(\X,0)+(0,-\dlY)$);
		\draw[signalH] ($(S1.east)+(0,-2*\dlY)$) to node[at end,right]{${\mathbf{Y}}_{+2}$} ($(S1.east)+(\X,0)+(0,-2*\dlY)$);
		
			
		\coordinate (O) at ($(9.0*\X,0)$);
		
		\node[square] (S2) at ($(O)+(S1)$) {};
		\node (T2) at ($(S2.north)+(0,\dlY)$) {With frequency coupling:};
		\draw[signalH] ($(S2.west)+(-\X,0)+(0,2*\dlY)$) to node[at start,left]{${\mathbf{U}}_{-2}$} ($(S2.west)+(0,2*\dlY)$);
		\draw[signalH] ($(S2.west)+(-\X,0)+(0,\dlY)$) to node[at start,left]{${\mathbf{U}}_{-1}$} ($(S2.west)+(0,\dlY)$);
		\draw[signalH] ($(S2.west)+(-\X,0)$) to node[at start,left]{${\mathbf{U}}_{0\phantom{-}}$} (S2.west);
		\draw[signalH] ($(S2.west)+(-\X,0)+(0,-\dlY)$) to node[at start,left]{${\mathbf{U}}_{+1}$} ($(S2.west)+(0,-\dlY)$);
		\draw[signalH] ($(S2.west)+(-\X,0)+(0,-2*\dlY)$) to node[at start,left]{${\mathbf{U}}_{+2}$} ($(S2.west)+(0,-2*\dlY)$);
		
		
		\draw[signalInC] ($(S2.west)+(0,2*\dlY)$) to ($(S2.east)+(0,\dlY)$);
		\draw[signalInC] ($(S2.west)+(0,\dlY)$) to ($(S2.east)+(0,0*\dlY)$);
		\draw[signalInC] ($(S2.west)+(0,\dlY)$) to ($(S2.east)+(0,2*\dlY)$);
		\draw[signalInC] ($(S2.west)$) to ($(S2.east)+(0,-\dlY)$);
		\draw[signalInC] ($(S2.west)$) to ($(S2.east)+(0,\dlY)$);
		\draw[signalInC] ($(S2.west)+(0,-\dlY)$) to ($(S2.east)+(0,-2*\dlY)$);
		\draw[signalInC] ($(S2.west)+(0,-\dlY)$) to ($(S2.east)+(0,-0*\dlY)$);
		\draw[signalInC] ($(S2.west)+(0,-2*\dlY)$) to ($(S2.east)+(0,-1*\dlY)$);
		
		\draw[signalInC2] ($(S2.west)+(0,2*\dlY)$) to ($(S2.east)+(0,0*\dlY)$);
		\draw[signalInC2] ($(S2.west)+(0,1*\dlY)$) to ($(S2.east)+(0,-1*\dlY)$);
		\draw[signalInC2] ($(S2.west)+(0,0*\dlY)$) to ($(S2.east)+(0,-2*\dlY)$);
		\draw[signalInC2] ($(S2.west)+(0,-2*\dlY)$) to ($(S2.east)+(0,0*\dlY)$);
		\draw[signalInC2] ($(S2.west)+(0,-1*\dlY)$) to ($(S2.east)+(0,1*\dlY)$);
		\draw[signalInC2] ($(S2.west)+(0,0*\dlY)$) to ($(S2.east)+(0,2*\dlY)$);
		
		\draw[signalInC3] ($(S2.west)+(0,2*\dlY)$) to ($(S2.east)+(0,-1*\dlY)$);
		\draw[signalInC3] ($(S2.west)+(0,1*\dlY)$) to ($(S2.east)+(0,-2*\dlY)$);
		\draw[signalInC3] ($(S2.west)+(0,-2*\dlY)$) to ($(S2.east)+(0,1*\dlY)$);
		\draw[signalInC3] ($(S2.west)+(0,-1*\dlY)$) to ($(S2.east)+(0,2*\dlY)$);
		
		\draw[signalInC4] ($(S2.west)+(0,-2*\dlY)$) to ($(S2.east)+(0,2*\dlY)$);
		\draw[signalInC4] ($(S2.west)+(0,2*\dlY)$) to ($(S2.east)+(0,-2*\dlY)$);
		
		\draw[signalIn] ($(S2.west)+(0,2*\dlY)$) to ($(S2.east)+(0,2*\dlY)$);
		\draw[signalIn] ($(S2.west)+(0,\dlY)$) to ($(S2.east)+(0,\dlY)$);
		\draw[signalIn] ($(S2.west)$) to (S2.east);
		\draw[signalIn] ($(S2.west)+(0,-\dlY)$) to ($(S2.east)+(0,-\dlY)$);
		\draw[signalIn] ($(S2.west)+(0,-2*\dlY)$) to ($(S2.east)+(0,-2*\dlY)$);
		
		\draw[signalH] ($(S2.east)+(0,2*\dlY)$) to node[at end,right]{${\mathbf{Y}}_{-2}$} ($(S2.east)+(\X,0)+(0,2*\dlY)$);
		\draw[signalH] ($(S2.east)+(0,\dlY)$) to node[at end,right]{${\mathbf{Y}}_{-1}$} ($(S2.east)+(\X,0)+(0,\dlY)$);
		\draw[signalH] (S2.east) to node[at end,right]{${\mathbf{Y}}_{0\phantom{-}}$} ($(S2.east)+\X*(1,0)$);
		\draw[signalH] ($(S2.east)+(0,-\dlY)$) to node[at end,right]{${\mathbf{Y}}_{+1}$} ($(S2.east)+(\X,0)+(0,-\dlY)$);
		\draw[signalH] ($(S2.east)+(0,-2*\dlY)$) to node[at end,right]{${\mathbf{Y}}_{+2}$} ($(S2.east)+(\X,0)+(0,-2*\dlY)$);
	
		
%
%
		
	\end{circuitikz}
}

%% file: Sections/Resource_Model.tex
\section{Harmonic State-Space Model of a \CIDER}
\label{sec:rsc-hss}



This chapter introduces the \HSS model of the individual \CIDER.
First, the theory of time-periodic signals and their Fourier series as well Toeplitz matrices are summarized in \cref{sec:rsc-hss:ltp}.
Second, the main hypotheses of the modelling framework introduced in \cite{jrn:2020:kettner-becker:HPF-1} are recalled in \cref{sec:rsc-hss:hyp}.
Third, the internal response of a generic \CIDER is recalled and a new formulation of the reference calculation is introduced in \cref{sec:rsc-hss:int-rep} and \cref{sec:rsc-hss:ref}.
Finally, the two are combined to form the \HSS model of the \CIDER in \cref{sec:rsc-hss:hss}.

\subsection{Primer on Time-Periodic Signals}
\label{sec:rsc-hss:ltp}

Harmonic analysis can be performed by means of \LTP systems theory~\cite{Ths:CSE:LTP:1991:Wereley}.
In this paper, all quantities are assumed to be time-periodic and \emph{Exponentially Modulated time-Periodic} (\EMP) w.r.t.~an underlying period $T$, which is the inverse of the fundamental frequency $f_{1}$ (i.e., $T=\frac{1}{f_{1}}$)%
.
Consider an exponentially modulated time-periodic vector $\mathbf{x}(t)$ and a time-periodic matrix $\mathbf{A}(t)$.
Any time-periodic signal can be represented by a Fourier series as
\begin{align}
	\mathbf{x}(t)
	&=	\sum\limits_{h\in\harmonics}\mathbf{X}_{h}\Exp{(s+j h 2\pi f_{1}) t}
	\label{eq:TP:vector}\\
	\mathbf{A}(t)
	&=  \sum\limits_{h\in\harmonics}\mathbf{A}_{h}\Exp{j h 2\pi f_{1} t}
	\label{eq:TP:matrix}
\end{align}
where $s \in \mathbb{C}$ is the Laplace operator and $\mathbf{X}_{h} \in \mathbb{C}$ is the complex Fourier coefficient of $\mathbf{x}(t)$ at the $h$-th harmonic of the fundamental frequency $f_{1}$, with $h\in\harmonics\subset\mathbb{Z}$%
\footnote{%
    Notably, $\harmonics\subset\mathbb{Z}$ is due to the representation of the positive and negative spectrum of the signal.
}.
Analogously, $\mathbf{A}_{h} \in \mathbb{C}$ is the complex Fourier coefficient of $\mathbf{A}(t)$ at the $h$-th harmonic.

The multiplication of two waveforms in time domain corresponds to the convolution of their spectra in frequency domain:
\begin{equation}
	\mathbf{A}(t)\mathbf{x}(t)
	\leftrightarrow \mathbf{A}(f)*\mathbf{X}(f)
	=               \hat{\mathbf{A}}\hat{\mathbf{X}}
	\label{eq:TP:time2freq}
\end{equation}
where $\hat{\mathbf{A}}$ is the Toeplitz matrix of the Fourier coefficients $\mathbf{A}_{h}$, and $\hat{\mathbf{X}}$ the column vector of the Fourier coefficients $\mathbf{X}_{h}$ \cite{Ths:CSE:LTP:1991:Wereley}
\begin{align}
	\hat{\mathbf{A}}	&:~	\hat{\mathbf{A}}_{\mathit{mk}}=\mathbf{A}_{h},~m,k\in\mathbb{N},~h=m-k\in\harmonics
	\label{eq:TP:constr}\\
	\hat{\mathbf{X}}	&=	\col_{h\in\harmonics}(\mathbf{X}_{h})
\end{align}
Unless the associated signals are band-limited, such matrices and vectors are of infinite size.
In practice, only the harmonics up to a certain maximum order $h_{max}$ are considered%
\footnote{%
	Standards for voltage and power quality typically account for harmonics up to order 20-25 (i.e., 1.0-1.5 kHz) \cite{Std:BSI-EN-50160:2000}.
}.
Hence, the said Toeplitz matrices and column vectors are of finite size.
In a Toeplitz matrix, the diagonal elements depict the direct link between identical frequencies, while the off-diagonal elements account for the coupling between different harmonics (i.e., similar to the illustration in \cref{fig:mf:ltp-coupling}).

\subsection{Underlying Hypotheses of the Modelling Framework}
\label{sec:rsc-hss:hyp}

A generic \CIDER can be of grid-forming or grid-following type.
A grid-forming \CIDER $\frm\in\Frm$ controls the voltage as a function of the current and 
a grid-following \CIDER $\flw\in\Flw$ controls the current as a function of the voltage at the point of connection to the grid.
To this end, within the modelling framework proposed in \cite{jrn:2020:kettner-becker:HPF-1,jrn:2020:kettner-becker:HPF-2} the resource are partitioned into grid-forming and grid-following type.

The studies performed in this paper are framed in the context of distribution system analysis, which allows to apply simplifications regarding the modelling of the power system components.
Specifically, in low-voltage distribution systems, the switching frequency of the \CIDER actuators is usually high, i.e., far beyond the frequency range that is of interest for harmonic analysis (i.e., up to 1-2~kHz)~\cite{Std:BSI-EN-50160:2000}.
Hence, in the frequency range of interest, the switching losses and high-frequency components due to the converter switching are negligible.
By consequence, the following hypothesis is made.
\begin{Hypothesis}\label{hyp:ph:act}
	In the context of harmonic analysis of power distribution systems, the actuator of the \CIDER[s] can be represented by an \emph{average model}~\cite{Dis:Peralta:2013}.
\end{Hypothesis}
If this assumption does not hold, the modelling framework does allow to include more complex models of the actuator thanks to its generality.
For instance, as discussed in Section~I of \cite{jrn:2022:becker}, switching effects can be modelled by a so-called \emph{Double Fourier Series} (\DFS), which involves the use of Bessel functions~\cite{Jrn:TIA:2009:McGrath}.
This aspect is beyond the scope of this paper.


\subsection{Internal Response}
\label{sec:rsc-hss:int-rep}
Within the framework proposed in \cite{jrn:2020:kettner-becker:HPF-1}, independent of the type of \CIDER, the so-called internal response of a \CIDER consists of the combination of the power hardware $\pwr$ and the control software $\ctrl$, which are interconnected through coordinate transformations.

As is introduced in detail in Section~IV-A of \cite{jrn:2020:kettner-becker:HPF-1}, all individual blocks of the \CIDER can be described by \LTP models or functions in time domain.
By transforming the \LTP models to harmonic domain using Fourier theory and Toeplitz matrices the blocks are represented in a tractable way, while accounting for the coupling between different harmonic frequencies (cf. Section~IV-B in \cite{jrn:2020:kettner-becker:HPF-1}).
Finally, the internal response of a \CIDER is described by a \HSS model that is derived as the closed-loop model between the \HSS models of the power hardware, control software and the coordinate transformations.
\begin{align}
	\Hat{\boldsymbol{\Psi}}\Hat{\mathbf{X}}
	&=      \Tilde{\mathbf{A}}\Hat{\mathbf{X}}
	+   \Tilde{\mathbf{E}}\Hat{\mathbf{W}}
	\label{eq:cider:proc:closed:hss}\\
	\Hat{\mathbf{Y}}
	&=      \Tilde{\mathbf{C}}\Hat{\mathbf{X}}
	+   \Tilde{\mathbf{F}}\Hat{\mathbf{W}}
	\label{eq:cider:meas:closed:hss}
\end{align}
where $\Hat{\mathbf{X}} = \col{(\Hat{\mathbf{X}}_\pwr,\Hat{\mathbf{X}}_\ctrl)}$ and equivalently for $\Hat{\mathbf{W}}$ and $\Hat{\mathbf{Y}}$ and the matrix $\hat{\boldsymbol{\Psi}}$ is given by
\begin{align}
    	\hat{\boldsymbol{\Psi}}
        &= s\cdot \diag(\mathbf{1})+j\hat{\boldsymbol{\Omega}}
	\label{eq:laplace}
\end{align}
with $\hat{\boldsymbol{\Omega}} = 2\pi f_1 \diag_{h\in\harmonics}(h\cdot\mathbf{1})$.
Note that the open-loop model of the \CIDER is characterized by block-diagonal matrices built from the power hardware and control software model. 
In contrast, the matrices of the closed-loop model in \eqref{eq:cider:proc:closed:hss}--\eqref{eq:cider:meas:closed:hss}, specifically $\Tilde{\mathbf{A}}$, $\Tilde{\mathbf{E}}$, $\Tilde{\mathbf{C}}$, and $\Tilde{\mathbf{F}}$, are densely populated.
For detailed derivations of the generic \CIDER model it is referred to Section~IV of \cite{jrn:2020:kettner-becker:HPF-1}.
In case nonlinearities are present in the power hardware or control software of the \CIDER, the model has to be linearized to derive the corresponding \LTP models.
As a consequence all matrices in \eqref{eq:cider:proc:closed:hss}--\eqref{eq:cider:meas:closed:hss} will be a function of the employed operating point.
This procedure is detailed in Section~II-C of~\cite{jrn:2022:becker}.

Since the state, disturbance and output vectors are defined as the column vectors of the respective quantities from the power hardware and the control software, the model introduced in \eqref{eq:cider:proc:closed:hss}--\eqref{eq:cider:meas:closed:hss} can be rewritten as follows:
\begin{align}
	\Hat{\boldsymbol{\Psi}}\Hat{\mathbf{X}}
	&=      \Tilde{\mathbf{A}}\Hat{\mathbf{X}}
	+   \Tilde{\mathbf{E}}_\pwr\Hat{\mathbf{W}}_\pwr
	+   \Tilde{\mathbf{E}}_\ctrl\Hat{\mathbf{W}}_\ctrl
	\label{eq:cider:proc:closed:hss:det}\\
	\Hat{\mathbf{Y}}
	&=      \Tilde{\mathbf{C}}\Hat{\mathbf{X}}
	+   \Tilde{\mathbf{F}}_\pwr\Hat{\mathbf{W}}_\pwr
	+   \Tilde{\mathbf{F}}_\ctrl\Hat{\mathbf{W}}_\ctrl
	\label{eq:cider:meas:closed:hss:det}
\end{align}
where $\Tilde{\mathbf{E}}_\pwr$ and $\Tilde{\mathbf{E}}_\ctrl$ represent the columns of $\Tilde{\mathbf{E}}$ associated with $\Hat{\mathbf{W}}_\pwr$ and $\Hat{\mathbf{W}}_\ctrl$, respectively, and analogously for $\Tilde{\mathbf{F}}_\pwr$ and $\Tilde{\mathbf{F}}_\ctrl$.

\begin{figure}[t]
    \centering
    \input{Figures/Resource_Model}
    \caption
    {%
        Block diagram of the generic model of \CIDER[s] in harmonic domain.
        The internal response is the closed-loop configuration between power hardware $\pwr$ and control software $\ctrl$.
        The reference calculation is represented by $\mathbf{\hat{R}}(\cdot,\cdot)$ and signals are subject to transformation matrices $\mathbf{\hat{T}}$ that represent changes of coordinate frames between power hardware and control software and circuit configurations between power hardware and grid~$\grd$.
    }
    \label{fig:CIDER:model}
\end{figure}
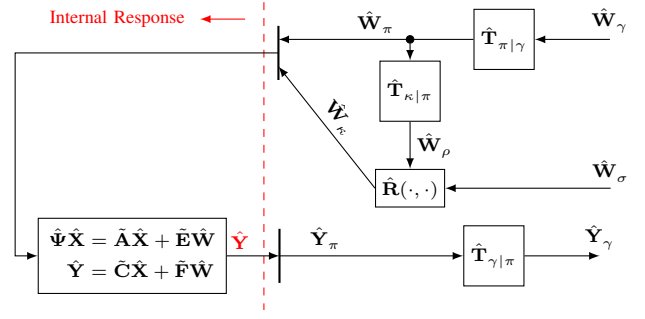


\subsection{Reference Calculation}
\label{sec:rsc-hss:ref}

The reference calculation $\refr$ is described by the function $\mathbf{r}(\cdot,\cdot)$ in time domain:
\begin{equation}
	\refr:\quad
	\mathbf{w}_{\ctrl}(t) = \mathbf{r}\left(\mathbf{w}_{\refr}(t),\mathbf{w}_{\spt}(t)\right)
	\label{eq:ref:time}
\end{equation}
where $\mathbf{w}_{\ctrl}$ is the disturbance of the control software, $\mathbf{w}_{\refr}$ is the grid disturbance described in the reference frame of the control software and $\mathbf{w}_{\spt}$ is the setpoint of the \CIDER.
It is important to note that $\mathbf{r}(\cdot,\cdot)$ need not be linear.
Specifically, for grid-following \CIDER[s] (i.e., with $\mathit{PQ}$ control), which compose the majority of resources in power grids, $\mathbf{r}(\cdot,\cdot)$ is usually nonlinear.
For grid-forming \CIDER[s] (i.e., with $\mathit{Vf}$ control), which are the minority of resources (typically only one), $\mathbf{r}(\cdot,\cdot)$ is usually linear.

One can either treat the nonlinearity in harmonic domain or perform a linearization of the reference calculation already in time domain (i.e., a small-signal model).
The first approach is introduced in Section~IV of \cite{jrn:2020:kettner-becker:HPF-1} and detailed for the case of a grid-following \CIDER in Section~III-C of \cite{jrn:2020:kettner-becker:HPF-2}.
Subsequently it is used for the derivation of the \CIDER response for the \HPF study.
The second approach, i.e., the small-signal model, is needed for deriving the \CIDER's \HSS model and its derivation is shown here below.
To this end, assume the following.
\begin{Hypothesis}\label{hyp:td:reference}
	The reference calculation $\mathbf{r}(\cdot,\cdot)$ is at least once differentiable w.r.t.~both $\mathbf{w}_{\refr}(t)$ and $\mathbf{w}_{\spt}(t)$ and can be represented by a small-signal model as:
	\begin{align}
		\begin{aligned}
			\mathbf{w}_{\ctrl}(t)
			\approx~\mathbf{\bar{w}}_{\ctrl}(t)
			&+	\mathbf{R}_{\refr}(t) \left(\mathbf{w}_{\refr}(t)-\bar{\mathbf{w}}_{\refr}(t)\right)\\
			&+	\mathbf{R}_{\spt}(t)\left(\mathbf{w}_{\spt}(t)-\bar{\mathbf{w}}_{\spt}(t)\right)
		\end{aligned}
		\label{eq:ref:multi}
	\end{align}
	where  $(\bar{\mathbf{w}}_{\refr}(t),\bar{\mathbf{w}}_{\spt}(t))$ describes the operating point w.r.t.~which the linearization is performed, $\mathbf{\bar{w}}_{\ctrl}(t) = \mathbf{r}\left(\mathbf{\bar{w}}_{\refr}(t),\mathbf{\bar{w}}_{\spt}(t)\right)$ represents a shift of the origin, and
	\begin{align}
		\mathbf{R}_{\refr}(t) &= \frac{\partial{\mathbf{r}\left(\bar{\mathbf{w}}_{\refr}(t),\bar{\mathbf{w}}_{\spt}(t)\right)}}{\partial{ \mathbf{w}_{\refr}(t)}} 
		  \label{eq:ref:multi:coef:refr}\\
		\mathbf{R}_{\spt}(t) &= \frac{\partial{ \mathbf{r}\left(\bar{\mathbf{w}}_{\refr}(t),\bar{\mathbf{w}}_{\spt}(t)\right)}}{\partial{\mathbf{w}_{\spt}(t)}} 
		\label{eq:ref:multi:coef:spt}
	\end{align}
\end{Hypothesis}
Notably, as opposed to conventional linearization in the context of \LTI systems, the operating point $(\bar{\mathbf{w}}_{\refr}(t),\bar{\mathbf{w}}_{\spt}(t))$ does not need to be constant, but can itself be a (known) time-periodic trajectory.

For the derivation of the \CIDER's \HSS model, the small-signal model of the reference calculation introduced in \eqref{eq:ref:multi} needs to be transformed to harmonic domain.
Recall that the matrices $\mathbf{R}_{\refr}(t)$ and $\mathbf{R}_{\spt}(t)$ in \eqref{eq:ref:multi:coef:refr}--\eqref{eq:ref:multi:coef:spt} are described by possibly nonlinear functions of the operating point and need to be approximated in the harmonic domain.
To this end, the following hypothesis has to hold.

\begin{Hypothesis}\label{hyp:reference:small-signal}
	There exist matrices $\hat{\mathbf{R}}_{\refr}$ and $\hat{\mathbf{R}}_{\spt}$ that approximate $\mathbf{R}_{\refr}(t)$ and $\mathbf{R}_{\spt}(t)$ of \eqref{eq:ref:multi} in the harmonic domain:
	\begin{align}
		\begin{aligned}
			\hat{\mathbf{W}}_{\ctrl}
			\approx	\hat{\bar{\mathbf{W}}}_{\ctrl}
			&+	\hat{\mathbf{R}}_{\refr}(\mathbf{\hat{\bar{W}}}_{\refr}, \mathbf{\hat{\bar{W}}}_{\spt}) \left[\hat{\mathbf{W}}_{\refr}-\mathbf{\hat{\bar{W}}}_{\refr}\right]\\
			&+	\hat{\mathbf{R}}_{\spt}(\mathbf{\hat{\bar{W}}}_{\refr}, \mathbf{\hat{\bar{W}}}_{\spt}) \left[\hat{\mathbf{W}}_{\spt}-\mathbf{\hat{\bar{W}}}_{\spt}\right]
		\end{aligned}	
		\label{eq:cider:hd:ref:small-signal}
	\end{align}	
	given the operating point composed of $\mathbf{\hat{\bar{W}}}_{\refr}$ and $\mathbf{\hat{\bar{W}}}_{\spt}$.
    $\hat{\bar{\mathbf{W}}}_{\ctrl}$ describes the shift of the origin to the operating point.
\end{Hypothesis}
\noindent
The Fourier coefficients of $\hat{\bar{\mathbf{W}}}_{\ctrl}$ can be computed beforehand using the possibly nonlinear approximation of the reference calculation as introduced in \cite{jrn:2020:kettner-becker:HPF-1}.
While this approximation might offer higher accuracy compared to the small-signal model described here, it is not suitable for the \HSS representation of the \CIDER.

Furthermore, rewrite the small-signal model of the reference calculation from \eqref{eq:cider:hd:ref:small-signal} as:
\begin{align}
	\hat{\mathbf{W}}_{\ctrl}
	=		\hat{\mathbf{R}}_{\opt} \hat{\mathbf{W}}_{\opt}
	+	\hat{\mathbf{R}}_{\refr} \Hat{\mathbf{T}}_{\ctrl|\pwr}\hat{\mathbf{W}}_{\pwr}
	+	\hat{\mathbf{R}}_{\spt} \hat{\mathbf{W}}_{\spt}
	\label{eq:cider:hd:ref:small-signal:hss}
\end{align}	
where 
\begin{align}
	\hat{\mathbf{R}}_{\opt}
	&=		\begin{bmatrix}
		\diag(\mathbf{1}) &-\hat{\mathbf{R}}_{\refr}\Hat{\mathbf{T}}_{\ctrl|\pwr}&-\hat{\mathbf{R}}_{\spt}
	\end{bmatrix}
	\\
	\hat{\mathbf{W}}_{\opt} 
	&= \col(\hat{\bar{\mathbf{W}}}_{\ctrl},\hat{\bar{\mathbf{W}}}_{\pwr},\hat{\bar{\mathbf{W}}}_{\spt})
\end{align}	
with $\diag(\mathbf{1})$ being an identity matrix of suitable size.
Notably, the coefficient matrices $\hat{\mathbf{R}}_{\opt}$, $	\hat{\mathbf{R}}_{\refr}$ and $\hat{\mathbf{R}}_{\spt}$ depend on the operating point $\hat{\mathbf{W}}_{\opt}$.
For the sake of clarity, this dependency is not restated for every instance of these matrices in the subsequent derivations.

\subsection{Combined Harmonic State-Space Model}
\label{sec:rsc-hss:hss}

For the \HSA, the grid response of the \CIDER is derived as the combination of the internal response in \eqref{eq:cider:proc:closed:hss:det}--\eqref{eq:cider:meas:closed:hss:det} and the small-signal representation of the reference calculation in \eqref{eq:cider:hd:ref:small-signal}, as well as the external transformations which are used for representing changes of circuit configurations between the \CIDER and the grid $\grd$. %
Note that the resulting \HSS model is capable of capturing internal characteristics related to the states of the \CIDER[s].

As introduced in Section~IV-B \cite{jrn:2020:kettner-becker:HPF-1} the external transformations in harmonic domain are given by:
\begin{align}
	\Hat{\mathbf{W}}_{\pwr}
	&=  \Hat{\mathbf{T}}_{\pwr|\grd}\Hat{\mathbf{W}}_{\grd}
	\label{eq:trafo:grd-to-pwr:freq}\\
	\Hat{\mathbf{Y}}_{\grd}
	&=  \Hat{\mathbf{T}}^{+}_{\grd|\pwr}\Hat{\mathbf{Y}}_{\pwr}
	\label{eq:trafo:pwr-to-grd:freq}
\end{align}

From this, one can extract the grid output $\Hat{\mathbf{Y}}_{\grd}$ from $\Hat{\mathbf{Y}}$ as follows:
\begin{align}
	\Hat{\mathbf{Y}}_{\grd}
	&=  \Hat{\mathbf{T}}^{+}_{\grd|\pwr}
	\begin{bmatrix}
		\diag(\mathbf{1}_\pwr) & \mathbf{0}_\ctrl
	\end{bmatrix}
	\Hat{\mathbf{Y}}
\end{align}
where $\diag(\mathbf{1}_\pwr)$ is the identity matrix and $\mathbf{0}_\ctrl$ the zero matrix whose sizes are compatible with $\Hat{\mathbf{Y}}_\pwr$ and $\Hat{\mathbf{Y}}_\ctrl$, respectively.

Combining \eqref{eq:cider:proc:closed:hss}--\eqref{eq:cider:meas:closed:hss} with \eqref{eq:cider:hd:ref:small-signal:hss} 
and adding the external transformations, leads to a state-space model describing the grid response of the \CIDER.
\begin{Definition}\label{def:mf:hss}
	The \emph{\HSS model} of the \CIDER describes the relation of the disturbances $\Hat{\mathbf{W}}_{\grd}$ and $\Hat{\mathbf{W}}_{\spt}$ w.r.t.~the grid output~$\hat{\mathbf{Y}}_\grd$ by the following equations:
	\begin{align}
		\Hat{\boldsymbol{\Psi}}\Hat{\mathbf{X}}
		&=      \Tilde{\mathbf{A}}\Hat{\mathbf{X}}
		+ 	 \Hat{\mathbf{E}}_\grd	\Hat{\mathbf{W}}_{\grd}
		+	 \Hat{\mathbf{E}}_{\spt} \hat{\mathbf{W}}_{\spt}
		+    \Hat{\mathbf{E}}_\opt \hat{\mathbf{W}}_{\opt}
		\label{eq:cider:proc:grid:hss}
		\\
		\Hat{\mathbf{Y}}_\grd
		&=      \Hat{\mathbf{C}}_\grd\Hat{\mathbf{X}}
		+ 	 \Hat{\mathbf{F}}_\grd	\Hat{\mathbf{W}}_{\grd}
		+	 \Hat{\mathbf{F}}_{\spt} \hat{\mathbf{W}}_{\spt}
		+    \Hat{\mathbf{F}}_\opt \hat{\mathbf{W}}_{\opt}
		\label{eq:cider:meas:grid:hss}
	\end{align}
	with coefficient matrices
	\begin{align}
		\Hat{\mathbf{E}}_\grd
		&=  
		\Tilde{\mathbf{E}}_\pwr\Hat{\mathbf{T}}_{\pwr|\grd} +	\Tilde{\mathbf{E}}_\ctrl\hat{\mathbf{R}}_{\refr} \Hat{\mathbf{T}}_{\ctrl|\pwr}\Hat{\mathbf{T}}_{\pwr|\grd}
		\\
	 	\Hat{\mathbf{E}}_\spt
		&=  \Tilde{\mathbf{E}}_\ctrl\hat{\mathbf{R}}_{\spt}
		\\
		 \Hat{\mathbf{E}}_\opt
		&=  \Tilde{\mathbf{E}}_\ctrl\hat{\mathbf{R}}_{\opt}
		\\
		\Hat{\mathbf{F}}_\grd
		&=  
	\Hat{\mathbf{T}}^{+}_{\grd|\pwr}\begin{bmatrix}\diag(\mathbf{1}_\pwr) & \mathbf{0}_\ctrl\end{bmatrix} \left(\Tilde{\mathbf{F}}_\pwr\Hat{\mathbf{T}}_{\pwr|\grd} +	\Tilde{\mathbf{F}}_\ctrl\hat{\mathbf{R}}_{\refr} \Hat{\mathbf{T}}_{\ctrl|\pwr}\Hat{\mathbf{T}}_{\pwr|\grd}\right)
		\\
		\Hat{\mathbf{F}}_\spt
		&=   
\Hat{\mathbf{T}}^{+}_{\grd|\pwr}	\begin{bmatrix}\diag(\mathbf{1}_\pwr) & \mathbf{0}_\ctrl\end{bmatrix} \Tilde{\mathbf{F}}_\ctrl\hat{\mathbf{R}}_{\spt}
		\\
		\Hat{\mathbf{F}}_\opt
		&=  
	\Hat{\mathbf{T}}^{+}_{\grd|\pwr}\begin{bmatrix}\diag(\mathbf{1}_\pwr) & \mathbf{0}_\ctrl\end{bmatrix}  \Tilde{\mathbf{F}}_\ctrl\hat{\mathbf{R}}_{\opt}
		\\
		\Hat{\mathbf{C}}_\grd
		&=  	 
	\Hat{\mathbf{T}}^{+}_{\grd|\pwr}\begin{bmatrix}\diag(\mathbf{1}_\pwr) & \mathbf{0}_\ctrl\end{bmatrix} \Tilde{\mathbf{C}}
	\end{align}
\end{Definition}
\noindent
Recall that the matrices describing the small-signal model of the reference calculation are functions of the operating point~$\hat{\mathbf{W}}_\opt$.
By consequence, the aforestated matrices of the \HSS model are dependent on this operating point, too.
Notably, from the \CIDER's \HSS the derivation of its \HTF is straight-forward.
Details regarding this can be found in Chapter~2.4.4 of \cite{ths:2024:becker}.


%% file: Figures/Resource_Model.tex
{

\tikzstyle{system}=[rectangle, draw=black, minimum width=1cm, minimum height=0.5cm, inner sep=0pt]
\tikzstyle{block}=[rectangle, draw=black, minimum size=0.5cm, inner sep=0pt]
\tikzstyle{dot}=[circle, draw=black, fill=black, minimum size=0.1cm, inner sep=0pt]

\tikzstyle{signal}=[-latex]

\definecolor{myRed}{rgb}{1 0 0}
\definecolor{myGreen}{rgb}{0 0 1}

\scriptsize

\begin{tikzpicture}

	\def\x{1}
	\def\y{0.9}
	
	
	
	\draw[dashed,draw=myRed] (1.55*\x,3.75*\y) to (1.55*\x,-0.8*\y);
	\draw[signal,draw=myRed] (1.3*\x,3.5*\y) to (0.7*\x,3.5*\y);
	\node[text=myRed] at (-0.4*\x,3.5*\y)
	{%
	    \begin{tabular}{c}
	        Internal Response
	    \end{tabular}
	};

	\node[rectangle,draw=black,minimum width=2.5cm,minimum height=1cm,inner sep=0pt](N) at (-0.2*\x,0) {%
        $
        \begin{aligned}
        	\Hat{\boldsymbol{\Psi}}\Hat{\mathbf{X}}
        	&=      \Tilde{\mathbf{A}}\Hat{\mathbf{X}}
        	+   \Tilde{\mathbf{E}}\Hat{\mathbf{W}}
            \\
        	\Hat{\mathbf{Y}}
        	&=      \Tilde{\mathbf{C}}\Hat{\mathbf{X}}
        	+   \Tilde{\mathbf{F}}\Hat{\mathbf{W}}
        \end{aligned}
        $
    };

    \coordinate (wnn) at ($(N.west)-(0.3*\x,0*\y)$);
	
	

	
	\coordinate (wng) at ($(wnn)+(3.5*\x,3.0*\y)$);
	\draw[thick] ($(wng)+(0*\x,-0.4*\y)$) to ($(wng)+(0*\x,0.4*\y)$);
	
	\draw[signal] (wng) 
	    to ($(wnn)+(0,3.0*\y)$)
	    to (wnn)
        to (N.west);
    
    \node[rectangle, draw=black, minimum size=8mm] (Tpg) at ($(wng)+(3.0*\x,0.2*\y)$) {$\hat{\mathbf{T}}_{\pwr|\grd}$};
    \node[dot] (wpn) at ($0.5*(wng)+0.5*(Tpg)+(0.25*\x,0.1*\y)$) {};
    \node[rectangle, draw=black, minimum size=5mm] (R) at ($(wng)+(1.75*\x,-2.0*\y)$) {$\hat{\mathbf{R}}(\cdot,\cdot)$};
    \node[rectangle, draw=black, minimum size=8mm] (Tkp) at ($0.4*(R.north)+0.6*(wpn)$) {$\hat{\mathbf{T}}_{\ctrl|\pwr}$};
    
    \draw[signal] (Tpg.west) to node[midway,above]{$\hat{\mathbf{W}}_{\pwr}$} ($(wng)+(0*\x,0.2*\y)$);
    
    \draw[signal] (wpn) to (Tkp.north);
    \draw[signal] (Tkp.south) to node[midway,right]{$\hat{\mathbf{W}}_{\refr}$}(R.north);
    \draw[signal] (R.west) to node[midway,above,sloped]{$\hat{\mathbf{W}}_{\ctrl}$} ($(wng)-(0*\x,0.2*\y)$);
	
	\coordinate (ws) at ($(R.east)+(2.2*\x,0)$);
	\draw[signal] (ws) to node[at start,above]{$\hat{\mathbf{W}}_{\spt}$} (R.east);
	\coordinate (wg) at ($(Tpg.east)+(1*\x,0)$);
	\draw[signal] (wg) to node[at start,above]{$\hat{\mathbf{W}}_{\grd}$} (Tpg.east);
	
    \node[rectangle, draw=black, minimum size=8mm] (Tgn) at ($(N.east)+(3.55*\x,0*\y)$) {$\hat{\mathbf{T}}_{\grd|\pwr}$};
    
    \coordinate (yng) at ($(N.east)+(0.7*\x,0*\y)$);
	\draw[thick] ($(yng)+(0*\x,-0.4*\y)$) to ($(yng)+(0*\x,0.4*\y)$);
	
	\draw[signal] (N.east) to node[near start,above,text=myRed]{$\hat{\mathbf{Y}}$} (yng);
	\draw[signal] (yng) to node[near start,above]{$\hat{\mathbf{Y}}_{\pwr}$} (Tgn.west);

	\coordinate (yg) at ($(Tgn.east)+(1*\x,0)$);
	\draw[signal] (Tgn.east) to node[at end,above]{$\hat{\mathbf{Y}}_{\grd}$} (yg);

\end{tikzpicture}

}

%% file: Sections/Grid_Model.tex
\section{Harmonic State-Space Model of the Grid}
\label{sec:grd-hss}

\begin{figure}[t]
	\centering
	\subfloat[]
	{%
		\centering
		\input{Figures/Grid_section_forming}
		\label{fig:Grid:model:forming}
	}
	
	\subfloat[]
	{%
		\centering
		\input{Figures/Grid_section_following}
		\label{fig:Grid:model:following}
	}
	\caption
	{%
		Lumped-element model of the grid shown at nodes with a grid-forming resource (\ref{fig:Grid:model:forming}) and a grid-following resource (\ref{fig:Grid:model:following}). 
		The electrical quantity controlled by the resource is highlighted in red.
	}
	\label{fig:Grid:model}
\end{figure}
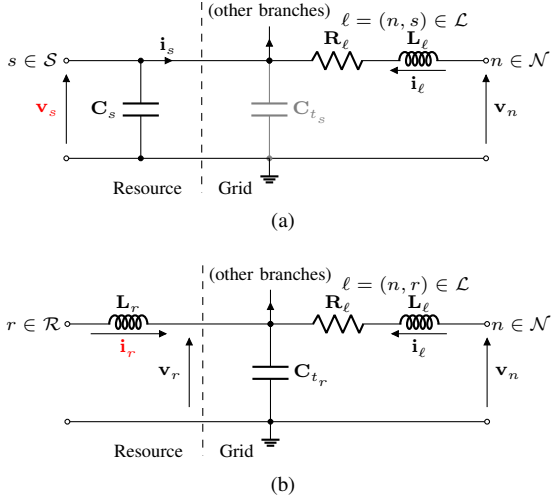

In this section, a state-space model of the grid is derived based on the grid topology given by all nodes $\nodes$ as well as the branches $\branches$ and shunts $\shunts$ of the line connecting them.
To this end, the branch and shunt elements are represented by a set of lumped-element models (see \cref{fig:Grid:model}) that are linear and passive (i.e., they contain no active elements, such as voltage or current sources).
In general, the state-space variables of the grid are given by i) the three-phase currents of the branch inductors and ii) the three-phase voltages of the shunt capacitors%
\footnote{%
	Within typical frequency ranges used in \HA (i.e., $f<10$~kHz) the lumped-element models of lines and transformers are such that the branches are inductive and shunts capacitive.
}.
Assume that the nodes are ordered as $\nodes=[\Frm,\Flw]$.
Then, the incidence matrix $\boldsymbol{\mathcal{A}}_{\branches|\nodes}$ that describes the topology of the grid can be expressed in terms of block matrices $\boldsymbol{\mathcal{A}}_{\branches|\nodes}=[\boldsymbol{\mathcal{A}}_{\branches|\Frm}~|~\boldsymbol{\mathcal{A}}_{\branches|\Flw}]$.

\subsection{Branches}
The dynamic equations related to the branch elements can be written as
\begin{align}
    \begin{aligned}
        \frac{d}{dt}\IT_{\branches}(t)
            =       - \LP_{\branches}^{-1}\RP_{\branches} \IT_{\branches}(t)
            &+   \LP_{\branches}^{-1} \boldsymbol{\mathcal{A}}_{\branches|\Frm} \VT_{\Frm}(t)\\
            &+   \LP_{\branches}^{-1} \boldsymbol{\mathcal{A}}_{\branches|\Flw} \VT_{\Flw}(t)
    \end{aligned}	
    \label{eq:grid:branches}
\end{align}
where $\VT_{\Frm}(t)=\col_{n\in\Frm}(\VT_{n}(t))$, $\VT_{\Flw}(t)=\col_{n\in\Flw}(\VT_{n}(t))$ are the three-phase voltages at nodes where grid-forming and grid-following \CIDER[s] are connected and $\IT_{\branches}(t)=\col_{\ell\in\branches}(\IT_{\ell}(t))$ are all three-phase branch currents.
The matrices in \eqref{eq:grid:branches} are block-diagonal and composed of the individual branch inductance and resistance matrices (i.e., $\LP_{\ell}$ and $\RP_{\ell}$).
More precisely, the matrices of all branches are given by $\LP_{\branches}=\diag_{\ell\in\branches}(\LP_{\ell})$ and $\RP_{\branches}=\diag_{\ell\in\branches}(\RP_{\ell})$, respectively.

\subsection{Shunts}
The shunt capacitors define the state-space variables associated with the nodes (i.e., the capacitor voltages).
They can be associated to i)~the line shunt parameters or ii)~the filter elements of the connected resource (i.e. the capacitance of the grid-forming resource).
To this end, the shunts are separated into two sets depending on which type of resource is connected at the corresponding node, i.e., grid-forming or grid-following resources.

\subsubsection{Nodes with Grid-Forming Resources}
At a node $\frm\in\Frm$, the resource controls the nodal voltage $\mathbf{v}_\frm$ over its output capacitor $\mathbf{C}_{\frm}$ (cf. \cref{fig:Grid:model:forming}). 
Since the shunt capacitor $\mathbf{C}_{t_\frm}$ is usually small compared to $\mathbf{C}_{\frm}$, it is neglected at grid-forming nodes.
This is usually true for low-voltage distribution grids. 
Thus, in view of \cref{sec:rsc-hss:hyp}, we can assume the following.
\begin{Hypothesis}\label{hyp:hss:grid:frm}
	In a low-voltage distribution system, the shunt capacitor $\mathbf{C}_{t_\frm}$ of a line connected to a grid-forming resource is small compared to $\mathbf{C}_{\frm}$.
\end{Hypothesis}
Then, the equation describing the nodes $\Frm\subset\nodes$, is given by
\begin{align}
	\IT_{\Frm}(t)
	&=      \boldsymbol{\mathcal{A}}_{\Frm|\branches} \IT_{\branches}(t)
\end{align}
with $\IT_{\Frm}(t)=\col_{n\in\Frm}(\IT_{n}(t))$ and $ \boldsymbol{\mathcal{A}}_{\Frm|\branches} =  \boldsymbol{\mathcal{A}}_{\branches|\Frm}^{\top}$.

\subsubsection{Nodes with Grid-Following Resources}
At a node $\flw\in\Flw$, the resource controls the current flowing through its grid-side filter inductance $\mathbf{L}_{\flw}$ and being injected into the grid (\cref{fig:Grid:model:following}).
The equations describing the nodes $\Flw\subset\nodes$, is given by 
\begin{align}
	\frac{d}{dt}\VT_{\Flw}(t)
	&=   \CP_{\shunts}^{-1} \boldsymbol{\mathcal{A}}_{\Flw|\branches} \IT_{\branches}(t)
	-  \CP_{\shunts}^{-1} \IT_{\Flw}(t)
\end{align}
where $ \boldsymbol{\mathcal{A}}_{\Flw|\branches} =  \boldsymbol{\mathcal{A}}_{\branches|\Flw}^{\top}$.


\subsection{Combined State-Space Model of the Grid}
In line with the previous explanations, the states, disturbances and outputs of the grid state-space model are defined as:
\begin{align}
	\XT_{\Grd}(t) &=
        \col(\IT_{\branches}(t), \VT_{\Flw}(t))
 \\
	\WT_{\Grd}(t) & =
        \col(\VT_{\Frm}(t), \IT_{\Flw}(t))
	\\
	\YT_{\Grd}(t)&  =
        \col(\IT_{\Frm}(t), \VT_{\Flw}(t))
\end{align}
It follows
\begin{align}
	\dot{\XT}_{\Grd}(t)
	&=      \mathbf{A}_{\Grd}(t)\XT_{\Grd}(t)
	+   \mathbf{E}_{\Grd}(t)\WT_{\Grd}(t)
	\label{eq:hsa:grd:prc}
	\\
	\YT_{\Grd}(t)
	&=      \mathbf{C}_{\Grd}(t)\XT_{\Grd}(t)
	+   \mathbf{F}_{\Grd}(t)\WT_{\Grd}(t)
	\label{eq:hsa:grd:meas}
\end{align}
with the matrices
\begin{align}
		\mathbf{A}_{\Grd}(t) &=\mathbf{A}_{\Grd,0}= \begin{bmatrix}
			- \LP_{\branches}^{-1}\RP_{\branches} & \LP_{\branches}^{-1}
			\boldsymbol{\mathcal{A}}_{\branches|\Flw}\\
			- \CP_{\shunts}^{-1} \boldsymbol{\mathcal{A}}_{\Flw|\branches} & \mathbf{0}
		\end{bmatrix}
	\\
		\mathbf{E}_{\Grd}(t) &=\mathbf{E}_{\Grd,0}= \begin{bmatrix}
			\LP_{\branches}^{-1}
			\boldsymbol{\mathcal{A}}_{\branches|\Frm} & \mathbf{0}\\
			\mathbf{0} &  \CP_{\shunts}^{-1}
		\end{bmatrix}
	\\
		\mathbf{C}_{\Grd}(t) &=\mathbf{C}_{\Grd,0}= \begin{bmatrix}
			\boldsymbol{\mathcal{A}}_{\Frm|\branches} & \mathbf{0}\\
			\mathbf{0} &  \mathbf{I}
		\end{bmatrix}
	\\
		\mathbf{F}_{\Grd}(t) &=\mathbf{F}_{\Grd,0}= \mathbf{0}
		\label{eq:hsa:grd:outputdist}
\end{align}

Transforming \eqref{eq:hsa:grd:prc}--\eqref{eq:hsa:grd:meas} to harmonic domain employing the Toeplitz transform yields the \HSS model of the grid:
\begin{align}
	\Hat{\boldsymbol{\Psi}}_{\Grd}\Hat{\mathbf{X}}_{\Grd}
	&=      \Hat{\mathbf{A}}_{\Grd}\Hat{\mathbf{X}}_{\Grd}
	+   \tilde{\mathbf{E}}_{\Grd}\tilde{\mathbf{W}}_{\Grd}
	\label{eq:grid:proc:closed}\\
	\tilde{\mathbf{Y}}_{\Grd}
	&=      \tilde{\mathbf{C}}_{\Grd}\Hat{\mathbf{X}}_{\Grd}
	+   \tilde{\mathbf{F}}_{\Grd}\tilde{\mathbf{W}}_{\Grd}
	\label{eq:grid:meas:closed}
\end{align}
with
\begin{alignat}{3}
	\Tilde{\mathbf{W}}_{\Grd} 
	&= \col_{h\in\harmonics}(\mathbf{W}_{\Grd,h}),~ &&\text{and}~\mathbf{w}_{\Grd}(t) &= \col(\mathbf{v}_{\Frm}(t),\mathbf{i}_{\Flw}(t))
	\label{eq:grid:disturbance}
	\\
	\Tilde{\mathbf{Y}}_{\Grd} 
	&= \col_{h\in\harmonics}(\mathbf{Y}_{\Grd,h}),~ &&\text{and}~\mathbf{y}_{\Grd}(t) &= \col(\mathbf{i}_{\Frm}(t),\mathbf{v}_{\Flw}(t))
	\label{eq:grid:output}
\end{alignat}   
where $\mathbf{W}_{\Grd,h}$ and $\mathbf{Y}_{\Grd,h}$ describe the Fourier coefficients of $\mathbf{w}_{\Grd}(t)$ and $\mathbf{y}_{\Grd}(t)$ at the harmonic $h\in\harmonics$, respectively.

%% file: Figures/Grid_section_forming.tex
{
	\tikzstyle{block}=[rectangle, draw=black,minimum size=1.1cm]
    \ctikzset{bipoles/length=0.8cm}
 
	\begin{circuitikz}
		
		\scriptsize
		
		
		\def\x{1.8}
		\def\y{1.3}
		
		
		
		\coordinate (NL) at (-1.5*\x,0);	
		\coordinate (NC) at (0,0);
		\coordinate (NR) at (1.6*\x,0);
		
		\coordinate (PL) at (-1.5*\x,\y);
		\coordinate (PC) at (0,\y);
		\coordinate (PR) at (1.6*\x,\y);
		
		
		
		\draw (PC)
		to[short] ($7/8*(PC)+1/8*(PR)$)
		to[resistor=$\mathbf{R}_{\ell}$] ($1/2*(PC)+1/2*(PR)$)
		to[inductor=$\mathbf{L}_{\ell}$,v_<=${\mathbf{i}_{\ell}}$] ($1/8*(PC)+7/8*(PR)$)
		to[short,-o] (PR);
		
		\draw (NR) to[short,o-] (NC);
		
		\node at ($(PR)+(-0.6*\x,0.4*\y)$) {$\ell=(n,\frm)\in\branches$};
		
		\node at ($(PR)+0.25*(\x,0)$) {$n\in\nodes$};
		\node at ($(NR)+0.25*(\x,0)$) {};
		
		
		
		\draw[gray] (PC)
		to[capacitor=$\mathbf{C}_{t_\frm}$,*-*,color=gray] (NC);
		
		
		
		\draw (NC)
		to[short,-] (NL)
		to[open,o-o,v^=\textcolor{red}{$\mathbf{v}_{\frm}$}] (PL)
		to[short,-*,i=${\mathbf{i}_\frm}$] (PC);
		\draw (NC) node[ground]{};
		
		\draw ($0.5*(NC)+0.5*(NL)-(0.2*\x,0)$) to[capacitor=$\mathbf{C}_{\frm}$,*-*] ($0.5*(PC)+0.5*(PL)-(0.2*\x,0)$);
		\draw (NR) to[open,v_=${\mathbf{v}_{n}}$] (PR);
		
		\node at ($(PL)-0.25*(\x,0)$) {$\frm\in\Frm$};
		
		
		\draw[dashed] ($(PC)+(-0.5*\x,0.6*\y)$) to ($(NC)+(-0.5*\x,-0.35*\y)$);
		\node at ($(NC)+(-0.9*\x,-0.3*\y)$) {Resource};
		\node at ($(NC)+(-0.25*\x,-0.3*\y)$) {Grid};
		
		
		
		\node (BO) at ($(PC)+0.5*(0,\y)$) {(other branches)};
		
		\draw[short,*-,i=$~$] (PC) to (BO.south);
		
	\end{circuitikz}
	
}

%% file: Figures/Grid_section_following.tex
{
	
	\tikzstyle{block}=[rectangle, draw=black,minimum size=1.1cm]
    \ctikzset{bipoles/length=0.8cm}
	
	\begin{circuitikz}
		
		\scriptsize
		
		
		\def\x{1.8}
		\def\y{1.3}
		
		
		
		\coordinate (NL) at (-1.5*\x,0);	
		\coordinate (NC) at (0,0);
		\coordinate (NR) at (1.6*\x,0);
		\coordinate (NS) at ($(NC)+(-0.5*\x,0*\y)$);
		
		\coordinate (PL) at (-1.5*\x,\y);
		\coordinate (PC) at (0,\y);
		\coordinate (PR) at (1.6*\x,\y);
		\coordinate (PS) at ($(PC)+(-0.5*\x,0*\y)$);
		
		
		
		\draw (PC)
		to[short] ($7/8*(PC)+1/8*(PR)$)
		to[resistor=$\mathbf{R}_{\ell}$] ($1/2*(PC)+1/2*(PR)$)
		to[inductor=$\mathbf{L}_{\ell}$,v_<=${\mathbf{i}_{\ell}}$] ($1/8*(PC)+7/8*(PR)$)
		to[short,-o] (PR);
		
		\draw (NR) to[short,o-] (NC);
		
		\node at ($(PR)+(-0.6*\x,0.4*\y)$) {$\ell=(n,\flw)\in\branches$};
		
		\node at ($(PR)+0.25*(\x,0)$) {$n\in\nodes$};
		\node at ($(NR)+0.25*(\x,0)$) {};
		
		
		
		\draw (PC)
		to[capacitor=$\mathbf{C}_{t_\flw}$,*-*] (NC);
		
		
		
		\draw (PL)
		to[inductor=$\mathbf{L}_{\flw}$,v_>=\textcolor{red}{${\mathbf{i}_{\flw}}$}] ($(PS)+(-0.1*\x,0)$)
		to[short] (PC);
		\draw (NC)
		to[short,-] (NL)
		to[open,o-o] (PL);
		\draw (NC) node[ground]{};
		
		\draw ($(NS)+(-0.1*\x,0)$)
		to [open,v^=${\mathbf{v}_{\flw}}$] ($(PS)+(-0.1*\x,0)$);
		
		\draw (NR) to[open,v_=${\mathbf{v}_{n}}$] (PR);
		
		\node at ($(PL)-0.25*(\x,0)$) {$\flw\in\Flw$};
		
		
		\draw[dashed] ($(PC)+(-0.5*\x,0.6*\y)$) to ($(NC)+(-0.5*\x,-0.35*\y)$);
		\node at ($(NC)+(-0.9*\x,-0.3*\y)$) {Resource};
		\node at ($(NC)+(-0.25*\x,-0.3*\y)$) {Grid};
		
		
		
		\node (BO) at ($(PC)+0.5*(0,\y)$) {(other branches)};
		
		\draw[short,*-,i=$~$] (PC) to (BO.south);
		
	\end{circuitikz}
	
}

%% file: Sections/System_Model.tex
\section{Harmonic State-Space Model of a Power System}
\label{sec:sys-hss}

In this section, the \HSS model of the entire power system is derived.
Through combination of the \HSS models of the resources and the grid, the open-loop \HSS model of the power system is derived in \cref{sec:hsa:sys-hss:ol}.
Subsequently, the closed-loop model is calculated following the electrical interconnection between the resources and the grid in \cref{sec:hsa:sys-hss:cl}.

\subsection{Open-Loop Model of the Power System}
\label{sec:hsa:sys-hss:ol}


Let $\rsc \in \Rsc$ be an generic CIDER (i.e., irrespective of the governing control law). 
In order to derive the \HSS of the entire system, one needs to combine the \HSS of the individual \CIDER[s] in \eqref{eq:cider:proc:grid:hss}--\eqref{eq:cider:meas:grid:hss} to the model of all resources.
\begin{alignat}{2}
	\Hat{\boldsymbol{\Psi}}_{\Rsc}\Hat{\mathbf{X}}_{\Rsc}
	&=      \Hat{\mathbf{A}}_{\Rsc}\Hat{\mathbf{X}}_{\Rsc}
            &&+ \sum_{j = \{\grd,\spt,\opt\}} \Hat{\mathbf{E}}_{\Rsc,j}	\Hat{\mathbf{W}}_{\Rsc,j}
	\label{eq:allrsc:proc:closed}
	\\
	\Hat{\mathbf{Y}}_{\Rsc,\grd}
	&=      \Hat{\mathbf{C}}_{\Rsc,\grd}\Hat{\mathbf{X}}_{\Rsc}
            &&+ \sum_{j = \{\grd,\spt,\opt\}} \Hat{\mathbf{F}}_{\Rsc,j}	\Hat{\mathbf{W}}_{\Rsc,j}
	\label{eq:allrsc:meas:closed}
\end{alignat}
where $\hat{\mathbf{X}}_{\Rsc}=	\col_{\rsc\in\Rsc}(\hat{\mathbf{X}}_{\rsc})$ and $\hat{\mathbf{A}}_{\Rsc}=  \diag_{\rsc\in\Rsc}(\hat{\mathbf{A}}_{\rsc})$, with $\hat{\mathbf{X}}_{\rsc}$ and $\hat{\mathbf{A}}_{\rsc}$ being the vectors and matrices of the \HSS of an individual \CIDER in \eqref{eq:cider:proc:grid:hss}--\eqref{eq:cider:meas:grid:hss}.
The remaining vectors (i.e., $\hat{\mathbf{Y}}_{\Rsc}$ etc.) and matrices (i.e., $\hat{\mathbf{E}}_{\Rsc}$ etc.) are defined analogously.
Without loss of generality, one can assume that the resources $\Rsc$ are ordered as $\Rsc = (\Frm,\Flw)$.
Then,
\begin{align}
	\Hat{\mathbf{W}}_{\Rsc,\grd} 
	&= \col(\Hat{\mathbf{I}}_{\Frm}, \Hat{\mathbf{V}}_{\Flw} )
	\label{eq:allrsc:disturbance}
	\\
	\Hat{\mathbf{Y}}_{\Rsc,\grd} 
	&= \col(\Hat{\mathbf{V}}_{\Frm} , \Hat{\mathbf{I}}_{\Flw} )
	\label{eq:allrsc:output}
\end{align}


As will be shown shortly, the \HSS model of all resources is interconnected with the \HSS model of the grid in closed-loop configuration.
Note that the Fourier coefficients which constitute $\Hat{\mathbf{W}}_{\Rsc,\grd}$ and $\Hat{\mathbf{Y}}_{\Rsc,\grd}$ in \eqref{eq:allrsc:disturbance}--\eqref{eq:allrsc:output} are grouped per node (i.e., the different harmonics associated with a given node form a block).
By contrast, $\Tilde{\mathbf{W}}_{\Grd} $ and $\Tilde{\mathbf{Y}}_{\Grd}$ in \eqref{eq:grid:disturbance}--\eqref{eq:grid:output} are grouped per harmonic order (i.e., the different nodal phasors associated with a given harmonic order form a block).
To combine the two models, a permutation from the grouping w.r.t.~the harmonics, here denoted by $\tilde{\mathbf{W}}$ and $\tilde{\mathbf{Y}}$, to the grouping w.r.t.~the nodes, denoted by $\hat{\mathbf{W}}$ and $\hat{\mathbf{Y}}$, and vice versa, is needed.
Then, the \HSS model of the grid can be rewritten as:
\begin{align}
	\Hat{\boldsymbol{\Psi}}_{\Grd}\Hat{\mathbf{X}}_{\Grd}
	&=      \Hat{\mathbf{A}}_{\Grd}\Hat{\mathbf{X}}_{\Grd}
	+   \Hat{\mathbf{E}}_{\Grd}\Hat{\mathbf{W}}_{\Grd}
	\label{eq:grid:proc:closed:reorderd}\\
	\Hat{\mathbf{Y}}_{\Grd}
	&=      \Hat{\mathbf{C}}_{\Grd}\Hat{\mathbf{X}}_{\Grd}
	+   \Hat{\mathbf{F}}_{\Grd}\Hat{\mathbf{W}}_{\Grd}
	\label{eq:grid:meas:closed:reorderd}
\end{align}
where a permutation of $\Tilde{\mathbf{W}}_{\Grd} $ and $\Tilde{\mathbf{Y}}_{\Grd}$ was performed and the corresponding matrices are updated accordingly.
Detailed definitions regarding this permutation matrix are given in \cite{jrn:2024:dipasquale}.
After applying the permutation the disturbance and output of the grid model are given by:
\begin{align}
	\Hat{\mathbf{W}}_{\Grd} 
	&= \col(\Hat{\mathbf{V}}_{\Frm}, \Hat{\mathbf{I}}_{\Flw} )
	\label{eq:grd:disturbance}
	\\
	\Hat{\mathbf{Y}}_{\Grd} 
	&= \col(\Hat{\mathbf{I}}_{\Frm} , \Hat{\mathbf{V}}_{\Flw} )
	\label{eq:grd:output}
\end{align}

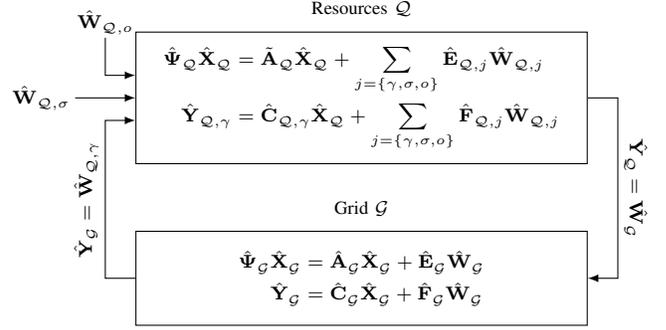
\begin{figure}[t]
	\centering
	\input{Figures/System_Model}
	\caption
	{%
		Block diagram of the generic power system.
	}
	\label{fig:hsa:sys:model}
\end{figure}

The interconnection of the resources and the grid is illustrated in \cref{fig:hsa:sys:model}.
As can be seen, the grid disturbance of the \HSS model of all resources is the output of the \HSS model of the grid, and vice versa.
To calculate the open-loop model of the system, the resrouces' \HSS model in \eqref{eq:allrsc:proc:closed}--\eqref{eq:allrsc:meas:closed} and the \HSS model of the grid \eqref{eq:grid:proc:closed:reorderd}--\eqref{eq:grid:meas:closed:reorderd} are combined into
\begin{align}
    \begin{aligned}
    	\Hat{\boldsymbol{\Psi}}_{\Prc}\Hat{\mathbf{X}}_{\Prc}
    	=      \hat{\mathbf{A}}_{\Prc}\Hat{\mathbf{X}}_{\Prc}
    	+   \hat{\mathbf{E}}_{\Prc,\grd}\Hat{\mathbf{W}}_{\Prc,\grd}
    	&+   \hat{\mathbf{E}}_{\Prc,\spt}\Hat{\mathbf{W}}_{\Rsc,\spt}\\
    	&+   \hat{\mathbf{E}}_{\Prc,\opt}\Hat{\mathbf{W}}_{\Rsc,\opt}
    \end{aligned}	
	\label{eq:process:proc:open}    \\
    \begin{aligned}
    	\Hat{\mathbf{Y}}_{\Prc}
    	=      \hat{\mathbf{C}}_{\Prc}\Hat{\mathbf{X}}_{\Prc}
    	+   \hat{\mathbf{F}}_{\Prc,\grd}\Hat{\mathbf{W}}_{\Prc,\grd}
    	&+   \hat{\mathbf{F}}_{\Prc,\spt}\Hat{\mathbf{W}}_{\Rsc,\spt}\\
    	&+   \hat{\mathbf{F}}_{\Prc,\opt}\Hat{\mathbf{W}}_{\Rsc,\opt}
    \end{aligned}	
	\label{eq:process:meas:open}
\end{align}
where
\begin{align}
	\hat{\mathbf{X}}_{\Prc}
	&=	\col(\hat{\mathbf{X}}_{\Rsc},\hat{\mathbf{X}}_{\Grd})
	\label{eq:process:var:state}
	\\
	\hat{\mathbf{W}}_{\Prc,\grd}
	&=	\col(\hat{\mathbf{W}}_{\Rsc,\grd},\hat{\mathbf{W}}_{\Grd})
	\label{eq:process:var:grid-disturbance}
	\\
	\hat{\mathbf{Y}}_{\Prc}
	&=	\col(\hat{\mathbf{Y}}_{\Rsc,\grd},\hat{\mathbf{Y}}_{\Grd})
	\label{eq:process:var:output}
\end{align}
as well as for the matrices
\begin{align}
	\hat{\mathbf{A}}_{\Prc}
	&=  \diag(\hat{\mathbf{A}}_{\Rsc},\hat{\mathbf{A}}_{\Grd})
	\label{eq:process:state}
	\\
	\hat{\mathbf{E}}_{\Prc,\grd}
	&=  \diag(\hat{\mathbf{E}}_{\Rsc,\grd},\hat{\mathbf{E}}_{\Grd})
	\label{eq:process:grid-disturbance}
\end{align}
and $\hat{\mathbf{\Psi}}_{\Prc}$, $\hat{\mathbf{C}}_{\Prc}$, and $\hat{\mathbf{F}}_{\Prc,\grd}$ are built analogously.

The matrices, associated with $\Hat{\mathbf{W}}_{\Rsc,\spt}$ and $\Hat{\mathbf{W}}_{\Rsc,\opt}$ are constructed as follows
\begin{align}
	\hat{\mathbf{E}}_{\Prc,\spt}
	&=  \col(\hat{\mathbf{E}}_{\Rsc,\spt},\mathbf{0})
	\label{eq:process:spt-disturbance}\\
	\hat{\mathbf{E}}_{\Prc,\opt}
	&=  \col(\hat{\mathbf{E}}_{\Rsc,\opt},\mathbf{0})
	\label{eq:process:opt-disturbance}
\end{align}
and  analogously for $\hat{\mathbf{F}}_{\Prc,\spt}$ and $\hat{\mathbf{F}}_{\Prc,\opt}$.


\subsection{Closed-Loop Model of the Power System}
\label{sec:hsa:sys-hss:cl}

From \eqref{eq:allrsc:disturbance}--\eqref{eq:allrsc:output} and \eqref{eq:grd:disturbance}--\eqref{eq:grd:output}, one can write the interconnection between the resources and the grid as:
\begin{align}
	\hat{\mathbf{W}}_{\Prc,\grd} = \Hat{\mathbf{J}}_{\Prc}\Hat{\mathbf{Y}}_{\Prc}
	\label{eq:process:trafo}
\end{align}
where
\begin{align}
	\Hat{\mathbf{J}}_{\Prc}
	&=
	\begin{bmatrix}
		\mathbf{0} & \diag(\mathbf{1})\\ 	 \diag(\mathbf{1}) &  \mathbf{0}
	\end{bmatrix}
\end{align}

One can interpret \eqref{eq:process:proc:open}--\eqref{eq:process:meas:open} as the open-loop model of the power system, and \eqref{eq:process:trafo} as the associated feedback control law.
In order to obtain the closed-loop model, substitute \eqref{eq:process:trafo} into \eqref{eq:process:proc:open}--\eqref{eq:process:meas:open} and solve for $\hat{\mathbf{X}}_{\Prc}$ and $\hat{\mathbf{Y}}_{\Prc}$:
\begin{align}
	\Hat{\boldsymbol{\Psi}}_{\Prc}\Hat{\mathbf{X}}_{\Prc}
	&=      \Tilde{\mathbf{A}}_{\Prc}\Hat{\mathbf{X}}_{\Prc}
	+   \Tilde{\mathbf{E}}_{\Prc,\spt}\Hat{\mathbf{W}}_{\Rsc,\spt}
	+   \Tilde{\mathbf{E}}_{\Prc,\opt}\Hat{\mathbf{W}}_{\Rsc,\opt}
	\label{eq:process:proc:closed}\\
	\Hat{\mathbf{Y}}_{\Prc}
	&=      \Tilde{\mathbf{C}}_{\Prc}\Hat{\mathbf{X}}_{\Prc}
	+   \Tilde{\mathbf{F}}_{\Prc,\spt}\Hat{\mathbf{W}}_{\Rsc,\spt}
	+   \Tilde{\mathbf{F}}_{\Prc,\opt}\Hat{\mathbf{W}}_{\Rsc,\opt}
	\label{eq:process:meas:closed}
\end{align}
where the matrices $\Tilde{\mathbf{A}}_{\Prc}$, $\Tilde{\mathbf{E}}_{\Prc,\spt}$, $\Tilde{\mathbf{E}}_{\Prc,\opt}$, $\Tilde{\mathbf{C}}_{\Prc}$, $\Tilde{\mathbf{F}}_{\Prc,\spt}$, and $\Tilde{\mathbf{F}}_{\Prc,\opt}$ are given by
\begin{align}
	\Tilde{\mathbf{A}}_{\Prc}
	&=  \Hat{\mathbf{A}}_{\Prc}+\Hat{\mathbf{E}}_{\Prc,\grd}(\diag(\mathbf{1})-\Hat{\mathbf{J}}_{\Prc}\Hat{\mathbf{F}}_{\Prc,\grd})^{-1}\Hat{\mathbf{J}}_{\Prc}\Hat{\mathbf{C}}_{\Prc}
	\label{eq:process:system}
	\\
	\Tilde{\mathbf{E}}_{\Prc,\spt}
	&=  \Hat{\mathbf{E}}_{\Prc,\spt}+\Hat{\mathbf{E}}_{\Prc,\grd}(\diag(\mathbf{1})-\Hat{\mathbf{J}}_{\Prc}\Hat{\mathbf{F}}_{\Prc,\grd})^{-1}\Hat{\mathbf{J}}_{\Prc}\Hat{\mathbf{F}}_{\Prc,\spt}
	\label{eq:process:input-spt}
	\\
	\Tilde{\mathbf{E}}_{\Prc,\opt}
	&=  \Hat{\mathbf{E}}_{\Prc,\opt}+\Hat{\mathbf{E}}_{\Prc,\grd}(\diag(\mathbf{1})-\Hat{\mathbf{J}}_{\Prc}\Hat{\mathbf{F}}_{\Prc,\grd})^{-1}\Hat{\mathbf{J}}_{\Prc}\Hat{\mathbf{F}}_{\Prc,\opt}
	\label{eq:process:input-opt}
	\\
	\Tilde{\mathbf{C}}_{\Prc}
	&=  (\diag(\mathbf{1})-\Hat{\mathbf{F}}_{\Prc,\grd}\Hat{\mathbf{J}}_{\Prc})^{-1}\Hat{\mathbf{C}}_{\Prc}
	\label{eq:process:output}
	\\
	\Tilde{\mathbf{F}}_{\Prc,\spt}
	&=  (\diag(\mathbf{1})-\Hat{\mathbf{F}}_{\Prc,\grd}\Hat{\mathbf{J}}_{\Prc})^{-1}\Hat{\mathbf{F}}_{\Prc,\spt}
	\label{eq:process:feedthrough-spt}
	\\
	\Tilde{\mathbf{F}}_{\Prc,\opt}
	&=  (\diag(\mathbf{1})-\Hat{\mathbf{F}}_{\Prc,\grd}\Hat{\mathbf{J}}_{\Prc})^{-1}\Hat{\mathbf{F}}_{\Prc,\opt}
	\label{eq:process:feedthrough-opt}
\end{align}
The matrices \eqref{eq:process:system}--\eqref{eq:process:feedthrough-opt} can only be computed if the inverses of $\diag(\mathbf{1})-\Hat{\mathbf{J}}_{\Prc}\Hat{\mathbf{F}}_{\Prc,\grd}$ and $\diag(\mathbf{1})-\Hat{\mathbf{F}}_{\Prc,\grd}\Hat{\mathbf{J}}_{\Prc}$ exist.
There is no general guarantee for this.
However, recall from \eqref{eq:hsa:grd:outputdist} that $\hat{\mathbf{F}}_{\Grd}=\mathbf{0}$, and hence, the aforementioned matrices are upper and lower block-triangular matrices, respectively.
The inverse of upper and lower block-triangular matrices does exist if the matrices on the diagonal are square and non-singular \cite{Jrn:Meyer:1970}.
Since the matrices on the daigaonal are identity matrices, one can deduce that the inverses of $\diag(\mathbf{1})-\Hat{\mathbf{J}}_{\Prc}\Hat{\mathbf{F}}_{\Prc,\grd}$ and $\diag(\mathbf{1})-\Hat{\mathbf{F}}_{\Prc,\grd}\Hat{\mathbf{J}}_{\Prc}$ exist.

Recall from \cref{sec:rsc-hss:hss} that the matrices of the \HSS of an individual \CIDER are a function of their operating point.
By consequence, the matrices of the \HSS 
of the system are also dependent on the operating point of all \CIDER[s], denoted by $\hat{\mathbf{W}}_{\Rsc,\opt}$ and $\hat{\mathbf{Y}}_{\Rsc,\opt}$.
For the sake of clarity, this dependency is not stated explicitly in the following.

%% file: Figures/System_Model.tex
{
	
\begin{tikzpicture}
	\tikzstyle{system}=[rectangle, draw=black, minimum width=6.0cm, minimum height=1.25cm, inner sep=0pt]
	\tikzstyle{block}=[rectangle, draw=black, minimum size=0.8cm, inner sep=0pt]
	\tikzstyle{dot}=[circle, draw=black, fill=black, minimum size=0.1cm, inner sep=0pt]
	\tikzstyle{signal}=[-latex]
	
	\scriptsize
	
	\def\dx{1.0}
	\def\dy{1.0}
	
	\node[system, minimum height=1.75cm] (P) at (-0.75*\dx,1.2*\dy) {
		$
		\begin{aligned}
			\Hat{\boldsymbol{\Psi}}_{\Rsc}\Hat{\mathbf{X}}_{\Rsc}
			&=      \Tilde{\mathbf{A}}_{\Rsc}\Hat{\mathbf{X}}_{\Rsc}
            + \sum_{j = \{\grd,\spt,\opt\}} \Hat{\mathbf{E}}_{\Rsc,j}	\Hat{\mathbf{W}}_{\Rsc,j}
			\\
			\Hat{\mathbf{Y}}_{\Rsc,\grd}
			&=      \Hat{\mathbf{C}}_{\Rsc,\grd}\Hat{\mathbf{X}}_{\Rsc}
            + \sum_{j = \{\grd,\spt,\opt\}} \Hat{\mathbf{F}}_{\Rsc,j}	\Hat{\mathbf{W}}_{\Rsc,j}
		\end{aligned}
		$
	};
	
	\node at ($(P.north)+(0,0.3*\dy)$) {Resources $\Rsc$};
	
	\node[system] (C) at (-0.75*\dx,-1.2*\dy) {
		$
		\begin{aligned}
			\Hat{\boldsymbol{\Psi}}_{\Grd}\Hat{\mathbf{X}}_{\Grd}
			&=      \Hat{\mathbf{A}}_{\Grd}\Hat{\mathbf{X}}_{\Grd}
			+   \Hat{\mathbf{E}}_{\Grd}\Hat{\mathbf{W}}_{\Grd}\\
			\Hat{\mathbf{Y}}_{\Grd}
			&=      \Hat{\mathbf{C}}_{\Grd}\Hat{\mathbf{X}}_{\Grd}
			+   \Hat{\mathbf{F}}_{\Grd}\Hat{\mathbf{W}}_{\Grd}
		\end{aligned}
		$
	};
	
	\node at ($(C.north)+(0,0.3*\dy)$) {Grid $\Grd$};
	
	\draw[signal] (P.east) to ($(P.east)+(0.4*\dx,0)$)
	to node[above,sloped,midway]{$\hat{\mathbf{Y}}_{\Rsc}=\hat{\mathbf{W}}_{\Grd}$} ($(C.east)+(0.4*\dx,0*\dy)$)
	to ($(C.east)+(0,0*\dy)$);
	
	\draw[signal] (C.west)
	to ($(C.west)-(0.4*\dx,0)$)
	to node[above,sloped,midway]{$\hat{\mathbf{Y}}_{\Grd}=\hat{\mathbf{W}}_{\Rsc,\grd}$} ($(P.west)-(0.4*\dx,0.3*\dy)$)
	to ($(P.west)-(0,0.3*\dy)$);
	
	\draw[signal] ($(P.west)+(-0.4*\dx,0.8*\dy)$)
	to ($(P.west)+(-0.4*\dx,0.3*\dy)$)
	to ($(P.west)+(0,0.3*\dy)$);
	\node at ($(P.west)+(-0.4*\dx,1.0*\dy)$) {$\hat{\mathbf{W}}_{\Rsc,\opt}$};
	\draw[signal] ($(P.west)+(-0.8*\dx,0*\dy)$)
	to ($(P.west)$);
	\node at ($(P.west)+(-1.25*\dx,0*\dy)$) {$\hat{\mathbf{W}}_{\Rsc,\spt}$};
	
\end{tikzpicture}
	
}

%% file: Sections/Operators.tex
\section{Operators for Harmonic Stability Assessment}
\label{sec:op-hsa}

This section gives a summary of the theory employed for the \HSA of the individual resources and the entire power system in Part~II of this paper.
First, the eigenvalue problem of a \HSS model and the associated eigenvectors are introduced in \cref{sec:op-hsa:eig}.
Second, the concept of sensitivity analysis w.r.t.~eigenvalue locations is given in \cref{sec:op-hsa:sens}, as well as a discussion of different types of eigenvalues in \cref{sec:op-hsa:class}.
Finally, an example of a set of eigenvalues and a discussion of the differences between \LTI and \HSS systems is given in \cref{sec:op-hsa:ex}.

\subsection{Eigenvalues and Eigenvectors}
\label{sec:op-hsa:eig}

Consider a generic \HSS model
\begin{align}
	\Hat{\boldsymbol{\Psi}}\Hat{\mathbf{X}}
	&=      \hat{\mathbf{A}}\Hat{\mathbf{X}}
	+   \hat{\mathbf{E}}\Hat{\mathbf{W}}
	\label{eq:hss:proc}\\
	\Hat{\mathbf{Y}}
	&=      \hat{\mathbf{C}}\Hat{\mathbf{X}}
	+   \hat{\mathbf{F}}\Hat{\mathbf{W}}
	\label{eq:hss:meas}
\end{align}
and its \HTF that describes the relation between the disturbance $\Hat{\mathbf{W}}$ and the output $\Hat{\mathbf{Y}}$:
\begin{align}
	\hat{\mathbf{Y}} = \hat{\mathbf{G}} \hat{\mathbf{W}}, \quad
    \hat{\mathbf{G}}
	= 	\hat{\mathbf{C}}(\Hat{\boldsymbol{\Psi}}-\hat{\mathbf{A}})^{-1}	\hat{\mathbf{E}}
	+	\hat{\mathbf{F}}
	\label{eq:hss:htf}
\end{align}
The harmonic stability of this \HSS model is determined through the poles of its \HTF.
More precisely, the poles of the system are the locations in the complex s-plane, where the \HTF is not analytical, i.e., its value is infinite \cite{Ths:CSE:LTP:1991:Wereley}.
Recall the composition of the matrix $\Hat{\boldsymbol{\Psi}}$ of a \HSS model as it was introduced in \eqref{eq:laplace}.
As a consequence, the poles of the \HTF are described by the eigenvalue problem associated with the matrix $\hat{\mathbf{A}}-j\hat{\boldsymbol{\Omega}}$:
\begin{align}
		\left(s\cdot \diag(\mathbf{1})-(\hat{\mathbf{A}}-j\hat{\boldsymbol{\Omega}})\right)\boldsymbol{\mathcal{V}}=\mathbf{0}
	\label{eq:hss:eig}
\end{align}
and $\boldsymbol{\mathcal{V}}$ is the matrix composed of the respective eigenvectors.

%
More precisely, each eigenvalue of the system matrix has an associated eigenvector.
One can analyse these vectors to understand the impact of a certain eigenvalue (also called mode) on a specific state variable, and vice versa.
As already mentioned in \cref{sec:soa}, numerous tools exist for this kind of analyses in the context of \LTI systems, i.e., participation factor analysis or modal analysis \cite{Bk:1994:Kundur} and generalisations to \LTP systems are proposed in \cite{Jrn:2020:Yang}.

\subsection{Sensitivity Analysis}
\label{sec:op-hsa:sens}

It is common practice to assess the sensitivity of the location of the eigenvalues w.r.t.~changes in the control parameters of the system.
To this end, the control parameters are varied, and the resulting changes of the locations of the eigenvalues assessed.
In this way, one can trace the so-called eigenvalue loci of the system.
Naturally, not all eigenvalues of the system are affected by those changes of the control parameters.
Some only change their location when the physical parameters of the system are varied, others do not change their location at all.

In order to assess such behaviour, the sets of eigenvalues calculated for each variation must be ordered identically. 
Unfortunately, state-of-the-art implementation, such as the \texttt{eig()} function of Matlab, does not ensure this. 
In this paper, the correct sorting is found using a \emph{Linear Assignment Problem} (\LAP).
The \LAP, given two sample sets, assigns pairs of samples, such that the total cost of assignment (described by a suitable cost function) is minimised.
In the context of sorting two sets of eigenvalues, the aim is to minimise the total distance between pairs of eigenvalues.
Let $\Lambda$ and $\tilde{\Lambda}$ be the two sets of eigenvalues to be sorted identically, and the cost function of assignment
\begin{align}
	\mathcal{C}: \Lambda \times \tilde{\Lambda} \rightarrow \mathbb{R}
\end{align}
Notably, the cost function $\mathcal{C}$ can be described by a matrix $\boldsymbol{\mathcal{C}}$:
\begin{align}
	\boldsymbol{\mathcal{C}}: (\boldsymbol{\mathcal{C}})_{ij} = |\lambda_i-\tilde{\lambda}_j|
\end{align}
Then, in order to solve the \LAP, find a bijection $f:\Lambda\longrightarrow\tilde{\Lambda}$, such that the total cost
\begin{align}
	\sum_{\lambda\in\Lambda}\boldsymbol{\mathcal{C}}_{[\lambda,f(\lambda)]}
\end{align}
is minimised \cite{Jrn:Duff:2001}.
In general, there is no guarantee that minimising this total distance will result in the correct sorting of the eigenvalues.
In particular, if the trajectories of two sets of eigenvalues intersect each other, pairs of eigenvalues may be associated wrongly in the vicinity of the intersection point.
In such situations, one can either include additional information about the characteristics of the eigenvalues (e.g., the eigenvectors) into the cost matrix or reduce the step size of the applied parameter change.

\subsection{Eigenvalue Classification and Spurious Eigenvalues}
\label{sec:op-hsa:class}

Three different types of eigenvalues are defined based on their impact on the \HSS model of a \CIDER (or any type of device).
To this end, the following naming is proposed: (i) the \emph{Control-Design Invariant} (\CDI), (ii) the  \emph{Control-Design Variant} (\CDV) and (iii) the \emph{Design Invariant} (\DI) eigenvalues.
Their definitions are given as:
\begin{Definition}\label{def:hsa:cdi}
	A \emph{\CDI} eigenvalue is an eigenvalue whose location remains fixed upon changing the parameters of the control system.
\end{Definition}
\begin{Definition}\label{def:hsa:cdv}
	The location of a \emph{\CDV} eigenvalue can be changed by modifying one or several of the parameters of the control system.
\end{Definition}
\begin{Definition}\label{def:hsa:di}
	A \emph{\DI} eigenvalue is an eigenvalue whose location remains fixed regardless of any parameter adjustments in the system.
\end{Definition}
As will be explained in Part~II of this paper, \DI eigenvalues of the \CIDER models occur due to the changes of the reference frame between power hardware (in this case: phase coordinates) and the control software (in this case: direct/quadrature components).

In a recent paper \cite{Jrn:Salis:2017}, which analyses eigenvalues of \LTP systems, the concept of 'spurious' eigenvalues is mentioned.
Recall that, when transforming \LTP models into \HSS models, Toeplitz matrices of infinite size would need to be used (i.e., in order to cover the entire, infinite spectrum).
In practice, those infinite Toeplitz matrices need to be truncated at a given maximum harmonic order.
Naturally, the truncation may introduce artefacts into the model, especially for the harmonics close to the boundaries of the model representation.
Therefore, eigenvalues, representing effects close to the maximum harmonic order might not be representative of the real system behaviour.
It is important to note that any tractable (i.e., finite) \HSS model inherently suffers from this effect due to the applied truncation.
For the case of the individual \CIDER[s], the \DI eigenvalues are one kind of the spurious eigenvalues.
The details of this analysis will be shown in Chapter~II.C of Part~II of this paper.
When combining several \HSS models (i.e., through closed-loop analysis), it becomes increasingly complex to discern which eigenvalues are genuine and which are spurious.
This challenge arises mainly because the mapping of the eigenvalues from an open-loop \HSS model to its closed-loop version is not straightforward in general.

\subsection{Example of Eigenvalues of an \HSS Model}
\label{sec:op-hsa:ex}

\begin{figure}[t]
	\centering
    \includegraphics[width=0.9\linewidth]{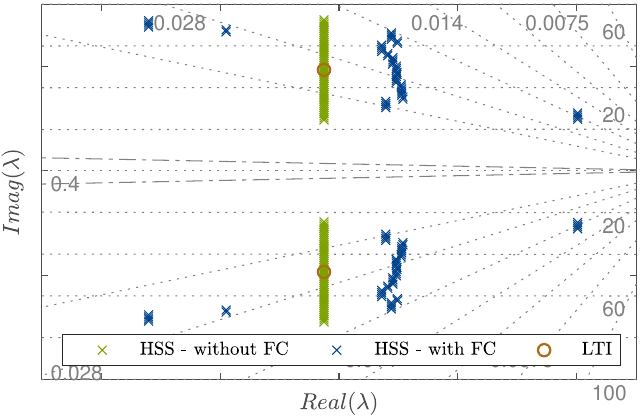}
	\caption
	{%
		Example of a mode of eigenvalues for an \LTI model, and a corresponding \HSS model with and without frequency coupling (FC) effect.
	}
	\label{fig:eigenvalue}
\end{figure}

\cref{fig:eigenvalue} compares the modes of eigenvalues obtained from an \HSS model with and without frequency coupling, as well as the eigenvalue pair of the corresponding \LTI system. 
The example shows one set of eigenvalues of the case study that will be introduced in Part~II of this paper.
Observe the translation of the eigenvalues of the \HSS models in direction of the imaginary axes.
This translation is caused by the matrix $\hat{\boldsymbol{\Omega}}$ in \eqref{eq:hss:eig} and relates to the different eigenvalues being associated to a specific frequency of the \HSS model.

In the absence of frequency coupling within the model, the system matrix of the \HSS model consists entirely of block-diagonal elements, with all off-diagonal elements being zero.
Therefore, the real parts of the eigenvalues of the \HSS model coincide with those of the \LTI model.
In this case the \HSS model provides no additional insight, since the represented dynamics are fully described by an \LTI model.

In contrast, for the model including frequency coupling, the system matrix will exhibit nonzero entries on the off-diagonal elements.
This results in a shift of the mode of eigenvalues as compared to the eigenvalues of the \LTI model.
In addition, the real parts of the translated eigenvalues do not match within the mode.
This stems from frequency-dependent operations (e.g., linearizations, modelling of time delays) performed in the harmonic domain.
Hence, in presence of frequency coupling in the model, the stability characteristics of the eigenvalues associated to different frequencies vary.

Note that the mode of eigenvalues of the \HSS model with frequency coupling exhibits eigenvalues with real parts differing strongly from those of the main mode.
These eigenvalues are to be considered spurious since they occur at the model boundary (i.e., at frequencies close to the maximum frequency taken into account).

%% file: Sections/Conclusion.tex
\section{Conclusions}
\label{sec:conclusion}

In this paper, a method for the \HSA of three-phase power grids with \CIDER[s] has been proposed.
The underlying modelling framework as it was proposed in \cite{jrn:2020:kettner-becker:HPF-1, jrn:2022:becker} is extended such that all power system components can be described by \HSS models.
The main changes compared to \cite{jrn:2020:kettner-becker:HPF-1, jrn:2022:becker} are i) that the reference calculation of the \CIDER[s] is represented by a small-signal model, and ii) the grid is described by a state-space model as opposed to the hybrid parameters.
These \HSS models of the power system components are combined to form the \HSS model of the entire power system, which can be employed for eigenvalue analysis in the context of \HSA.
Furthermore, a set of operators for the \HSA are introduced, such as the sensitivity analysis of the eigenvalue loci w.r.t. control parameters of the \CIDER[s], as well as a classification of the eigenvalues.
Finally, an illustration of the differences between the \HSA of a model with and without frequency coupling is discussed.

%% file: Main.bbl
\begin{thebibliography}{10}
\providecommand{\url}[1]{#1}
\csname url@samestyle\endcsname
\providecommand{\newblock}{\relax}
\providecommand{\bibinfo}[2]{#2}
\providecommand{\BIBentrySTDinterwordspacing}{\spaceskip=0pt\relax}
\providecommand{\BIBentryALTinterwordstretchfactor}{4}
\providecommand{\BIBentryALTinterwordspacing}{\spaceskip=\fontdimen2\font plus
\BIBentryALTinterwordstretchfactor\fontdimen3\font minus \fontdimen4\font\relax}
\providecommand{\BIBforeignlanguage}[2]{{%
\expandafter\ifx\csname l@#1\endcsname\relax
\typeout{** WARNING: IEEEtran.bst: No hyphenation pattern has been}%
\typeout{** loaded for the language `#1'. Using the pattern for}%
\typeout{** the default language instead.}%
\else
\language=\csname l@#1\endcsname
\fi
#2}}
\providecommand{\BIBdecl}{\relax}
\BIBdecl

\bibitem{Bk:1994:Kundur}
P.~Kundur, \emph{Power system stability and control}.\hskip 1em plus 0.5em minus 0.4em\relax McGraw-Hill Education, 1994.

\bibitem{Jrn:Paolone:2020}
M.~Paolone, T.~Gaunt, X.~Guillaud, M.~Liserre, S.~Meliopoulos, A.~Monti, T.~Van~Cutsem, V.~Vittal, and C.~Vournas, ``Fundamentals of power systems modelling in the presence of converter-interfaced generation,'' \emph{Electric Power Systems Research}, vol. 189, p. 106811, 2020.

\bibitem{Rep:PSE:SA:2018:Canizares}
C.~A. Ca{\~n}izares \emph{et~al.}, ``Microgrid stability definitions, analysis, and modeling,'' IEEE PES, Tech. Rep. PES-TR66, 2018.

\bibitem{Rep:PSE:SA:2020:Hatziargyriou}
N.~Hatziargyriou \emph{et~al.}, ``Stability definitions and characterization of dynamic behavior in systems with high penetration of power electronic interfaced technologies,'' IEEE PES, Tech. Rep. PES-TR77, 2020.

\bibitem{Jrn:PSE:PEC:2004:Enslin}
J.~H.~R. Enslin and P.~J.~M. Heskes, ``Harmonic interaction between a large number of distributed power inverters and the distribution network,'' \emph{IEEE Trans. Power Electron.}, vol.~19, no.~6, pp. 1586--1593, 2004.

\bibitem{Jrn:Wang:2018}
X.~Wang and F.~Blaabjerg, ``Harmonic stability in power electronic-based power systems: Concept, modeling, and analysis,'' \emph{IEEE Trans. Smart Grid}, vol.~10, no.~3, pp. 2858--2870, 2018.

\bibitem{Ths:CSE:LTP:1991:Wereley}
N.~M. Wereley, ``Analysis and control of linear periodically time-varying systems,'' Ph.D. dissertation, MIT, Cambridge, MA, USA, 1991.

\bibitem{Rep:PSE:SA:2004:Kundur}
P.~S. Kundur \emph{et~al.}, ``Definition and classification of power system stability,'' \emph{IEEE Trans. Power Syst.}, vol.~19, no.~3, pp. 1387--1401, 2004.

\bibitem{jrn:2020:kettner-becker:HPF-1}
A.~M. Kettner, L.~Reyes-Chamorro, J.~K.~M. Becker, Z.~Zou, M.~Liserre, and M.~Paolone, ``Harmonic power-flow study of polyphase grids with converter-interfaced distributed energy resources—part i: Modeling framework and algorithm,'' \emph{IEEE Trans. Smart Grid}, vol.~13, no.~1, pp. 458--469, 2021.

\bibitem{jrn:2020:kettner-becker:HPF-2}
J.~K.~M. Becker, A.~M. Kettner, L.~Reyes-Chamorro, Z.~Zou, M.~Liserre, and M.~Paolone, ``Harmonic power-flow study of polyphase grids with converter-interfaced distributed energy resources—part ii: Model library and validation,'' \emph{IEEE Trans. Smart Grid}, vol.~13, no.~1, pp. 470--481, 2021.

\bibitem{jrn:2022:becker}
J.~K.~M. Becker, A.~M. Kettner, Y.~Zuo, F.~Cecati, S.~Pugliese, M.~Liserre, and M.~Paolone, ``Modelling of ac/dc interactions of converter-interfaced resources for harmonic power-flow studies in microgrids,'' \emph{IEEE Trans. on Smart Grid}, vol.~14, no.~3, pp. 2096--2110, 2022.

\bibitem{Rep:2014:CIGRE}
K.~Strunz \emph{et~al.}, ``Benchmark systems for network integration of renewable and distributed energy resources,'' CIGR{\'E}, Paris, IDF, FR, Tech. Rep. 575, 2014.

\bibitem{Bk:PSE:SSA:1997:Arrillaga}
J.~Arrillaga, B.~C. Smith, N.~R. Watson, and A.~R. Wood, \emph{Power System Harmonic Analysis}.\hskip 1em plus 0.5em minus 0.4em\relax Hoboken, NJ, USA: Wiley, 1997.

\bibitem{Jrn:2017:Kwon}
J.~Kwon, X.~Wang, F.~Blaabjerg, C.~L. Bak, A.~R. Wood, and N.~R. Watson, ``Linearized modeling methods of ac--dc converters for an accurate frequency response,'' \emph{IEEE Journal of Emerging and Selected Topics in Power Electronics}, vol.~5, no.~4, pp. 1526--1541, 2017.

\bibitem{Jrn:Park:1929}
R.~H. Park, ``Two-reaction theory of synchronous machines,'' \emph{Trans. AIEE}, vol.~48, no.~3, pp. 716--727, 7 1929.

\bibitem{Jrn:Wang:2014}
X.~Wang, F.~Blaabjerg, and W.~Wu, ``Modeling and analysis of harmonic stability in an ac power-electronics-based power system,'' \emph{IEEE Trans. on power electronics}, vol.~29, no.~12, pp. 6421--6432, 2014.

\bibitem{Jrn:Yoon:2016}
C.~Yoon, H.~Bai, R.~N. Beres, X.~Wang, C.~L. Bak, and F.~Blaabjerg, ``Harmonic stability assessment for multiparalleled, grid-connected inverters,'' \emph{IEEE Trans. on Sustainable Energy}, vol.~7, no.~4, pp. 1388--1397, 2016.

\bibitem{Jrn:Kwon:2016}
J.~Kwon, X.~Wang, F.~Blaabjerg, C.~L. Bak, V.-S. Sularea, and C.~Busca, ``Harmonic interaction analysis in a grid-connected converter using harmonic state-space (hss) modeling,'' \emph{IEEE Trans. Power Electron.}, vol.~32, no.~9, pp. 6823--6835, 2016.

\bibitem{Jrn:Kwon:2016-1}
J.~B. Kwon, X.~Wang, F.~Blaabjerg, C.~L. Bak, A.~R. Wood, and N.~R. Watson, ``Harmonic instability analysis of a single-phase grid-connected converter using a harmonic state-space modeling method,'' \emph{IEEE Trans. on Industry Applications}, vol.~52, no.~5, pp. 4188--4200, 2016.

\bibitem{Jrn:2022:Yang}
H.~Yang, M.~Eggers, P.~Teske, and S.~Dieckerhoff, ``Comparative stability analysis and improvement of grid-following converters using novel interpretation of linear time-periodic theory,'' \emph{IEEE Journal of Emerging and Selected Topics in Power Electronics}, vol.~10, no.~6, pp. 7049--7061, 2022.

\bibitem{Jrn:Golestan:2021}
S.~Golestan, J.~M. Guerrero, A.~M. Abusorrah, J.~C. Vasquez, and Y.~Al-Turki, ``Ltp modeling and stability assessment of multiple second-order generalized integrator-based signal processing/synchronization algorithms and their close variants,'' \emph{IEEE Trans. on Power Electronics}, vol.~37, no.~5, pp. 5062--5077, 2021.

\bibitem{Cnf:Salis:2016}
V.~Salis, A.~Costabeber, P.~Zanchetta, and S.~Cox, ``Stability analysis of single-phase grid-feeding inverters with pll using harmonic linearisation and linear time periodic (ltp) theory,'' in \emph{2016 IEEE 17th Workshop on Control and Modeling for Power Electronics (COMPEL)}.\hskip 1em plus 0.5em minus 0.4em\relax IEEE, 2016, pp. 1--7.

\bibitem{Cnf:Stankovic:2000}
A.~Stankovic, P.~Mattavelli, V.~Caliskan, and G.~Verghese, ``Modeling and analysis of facts devices with dynamic phasors,'' in \emph{2000 IEEE Power Engineering Society Winter Meeting. Conference Proceedings (Cat. No.00CH37077)}, vol.~2, 2000, pp. 1440--1446 vol.2.

\bibitem{Cnf:Demiray:2008}
T.~Demiray, G.~Andersson, and L.~Busarello, ``Evaluation study for the simulation of power system transients using dynamic phasor models,'' in \emph{2008 IEEE/PES Transmission and Distribution Conference and Exposition: Latin America}, 2008, pp. 1--6.

\bibitem{Jrn:DeRua:2020}
P.~De~Rua, {\"O}.~C. Sakinci, and J.~Beerten, ``Comparative study of dynamic phasor and harmonic state-space modeling for small-signal stability analysis,'' \emph{Electric Power Systems Research}, vol. 189, p. 106626, 2020.

\bibitem{Cnf:Hall:1990}
S.~R. Hall and N.~M. Wereley, ``Generalized nyquist stability criterion for linear time periodic systems,'' in \emph{1990 American Control Conference}.\hskip 1em plus 0.5em minus 0.4em\relax IEEE, 1990, pp. 1518--1525.

\bibitem{Jrn:Salis:2017}
V.~Salis, A.~Costabeber, S.~M. Cox, P.~Zanchetta, and A.~Formentini, ``Stability boundary analysis in single-phase grid-connected inverters with pll by ltp theory,'' \emph{IEEE Trans. on Power Electronics}, vol.~33, no.~5, pp. 4023--4036, 2017.

\bibitem{Cnf:Kwon:2016}
J.~Kwon, X.~Wang, F.~Blaabjerg, and C.~L. Bak, ``Comparison of lti and ltp models for stability analysis of grid converters,'' in \emph{2016 IEEE 17th Workshop on Control and Modeling for Power Electronics (COMPEL)}.\hskip 1em plus 0.5em minus 0.4em\relax IEEE, 2016, pp. 1--8.

\bibitem{Jrn:Peng:2019}
Y.~Peng, Z.~Shuai, X.~Liu, Z.~Li, J.~M. Guerrero, and Z.~J. Shen, ``Modeling and stability analysis of inverter-based microgrid under harmonic conditions,'' \emph{IEEE Trans. on Smart Grid}, vol.~11, no.~2, pp. 1330--1342, 2019.

\bibitem{Jrn:2020:Yang}
H.~Yang, H.~Just, M.~Eggers, and S.~Dieckerhoff, ``Linear time-periodic theory-based modeling and stability analysis of voltage-source converters,'' \emph{IEEE Journal of Emerging and Selected Topics in Power Electronics}, vol.~9, no.~3, pp. 3517--3529, 2020.

\bibitem{Std:BSI-EN-50160:2000}
``Voltage characteristics of electricity supplied by public distribution networks,'' British Standards Institution, London, UK, Std. BS-EN-50160:2000, 2000.

\bibitem{Dis:Peralta:2013}
J.~Peralta~Rodriguez, ``Dynamic averaged models of vsc-based hvdc systems for electromagnetic transient programs,'' Ph.D. dissertation, {\'E}cole Polytechnique de Montr{\'e}al, 2013.

\bibitem{Jrn:TIA:2009:McGrath}
B.~P. McGrath and D.~G. Holmes, ``A general analytical method for calculating inverter dc-link current harmonics,'' \emph{IEEE Trans. Ind. Appl.}, vol.~45, no.~5, pp. 1851--1859, 2009.

\bibitem{ths:2024:becker}
J.~K.~M. Becker, ``Unified harmonic power-flow and stability analysis of power grids with converter-interfaced distributed energy resources,'' Ph.D. dissertation, EPFL, 2024.

\bibitem{jrn:2024:dipasquale}
A.~D. Pasquale, J.~K.~M. Becker, A.~M. Kettner, and M.~Paolone, ``Ensuring solution uniqueness in fixed-point-based harmonic power flow analysis with converter-interfaced resources: Ex-post conditions,'' 2024.

\bibitem{Jrn:Meyer:1970}
C.~D. Meyer, Jr, ``Generalized inverses of block triangular matrices,'' \emph{SIAM Journal on Applied Mathematics}, vol.~19, no.~4, pp. 741--750, 1970.

\bibitem{Jrn:Duff:2001}
I.~S. Duff and J.~Koster, ``On algorithms for permuting large entries to the diagonal of a sparse matrix,'' \emph{SIAM Journal on Matrix Analysis and Applications}, vol.~22, no.~4, pp. 973--996, 2001.

\end{thebibliography}
